\documentclass[prd,aps,10pt,twocolumn,floatfix,nofootinbib,superscriptaddress]{revtex4-1}
\usepackage{lipsum}
\usepackage{textcomp}
\usepackage{amsmath}
\usepackage{amsfonts}
\usepackage{amssymb}
\usepackage{graphicx}
\usepackage{hyperref}
\usepackage{subfigure}
\usepackage[usenames, dvipsnames]{color}
\usepackage{comment}

\begin{document}

\title{Magnetosphere of an orbiting neutron star} 

\author{Federico Carrasco}
\email{federico.carrasco@aei.mpg.de}
\affiliation{Max Planck Institute for Gravitational Physics (Albert Einstein Institute), 14476 Potsdam, Germany}
\author{Masaru Shibata}
\email{masaru.shibata@aei.mpg.de }
\affiliation{Max Planck Institute for Gravitational Physics (Albert Einstein Institute), 14476 Potsdam, Germany}
\affiliation{Center for Gravitational Physics, Yukawa Institute for Theoretical Physics, Kyoto University, 606-8502 Kyoto, Japan}

\date{\today}

\begin{abstract}
 
We conduct force-free simulations of a single neutron star undergoing orbital motion in flat spacetime, mimicking the trajectory of the star about the center of mass on a compact binary system. Our attention is focused on the kinetic energy being extracted from the orbit by the acceleration of the magnetic dipole moment of the neutron star, and particularly, on how this energy gets distributed within its surrounding magnetosphere. A detailed study of the resulting magnetospheric configurations in our setting is presented, incorporating as well the effects due to neutron star spin and the misalignment of the magnetic and orbital axes. We find many features resembling those of pulsar magnetospheres for the orbiting neutron star --even in the absence of spin--, being of particular interest the development of a spiral current sheet that extends beyond the light cylinder. Then, we use recent advances in pulsar theory to estimate electromagnetic emissions produced at the reconnection regions of such current sheets. 

\end{abstract}


\maketitle

\section{Introduction}

A new era of multi-messenger astronomy has started with the detection of gravitational waves (GW) from a binary neutron star merger (GW170817) by advanced LIGO/advanced Virgo~\cite{abbott17a,abbott17b}, followed by broadband electromagnetic (EM) observations \cite{abbott17c}. 
Binary systems involving a neutron star (NS) are the most likely sources for such simultaneous detections of GW and EM signals. In this context, 
EM emissions from the relatively cleaner environment preceding the merger could provide crucial information about the merger process, sky localization of the source, and the physical parameters of the system, which cannot be accurately obtained only by the gravitational-wave observation. 

The NS in the binary is generally expected to possess an approximately dipolar magnetic field, and they are likely to be surrounded by a force-free magnetosphere. The source for the precursor EM counterparts comes fundamentally from the orbital and rotational energy of the binary and its individual constituents. This kinetic energy is first electromagnetically extracted from each compact object, by means of the surrounding plasma, and later re-processed within the magnetosphere to produce the observable emissions. Starting from the pioneering work of Goldreich \& Julian \cite{goldreich} and Blandford \& Znajek \cite{Blandford}, the mechanisms to tap rotational energy from a compact object (immersed in a tenuous plasma) were unveiled: while an NS admits a classical electrodynamics interpretation as a \textit{Faraday disk}, for a spinning black hole (BH) the energy is instead extracted in a form of generalized Penrose process, widely known as the \textit{Blandford-Znajek mechanism}.
For the orbital motion, on the other hand, there are two such extraction mechanisms identified. 
The first one is associated with a purely classical effect, produced by the acceleration of a magnetic (e.g.,~dipole) moment \cite{landau1975}.  
The EM energy loss rate from the motion of a single NS with a dipole magnetic field, as well as for two NSs in a binary, has been estimated (assuming vacuum) in \cite{ioka2000}. 
The second mechanism is known as \textit{unipolar induction}, which essentially captures the transfer of linear momentum into EM energy, due to the motion of a conductor (or compact object) through a magnetized medium.
This effect has been studied long time ago for moving conductors such as satellites (see, e.g.,~\cite{1965drag, goldreich1969}), later extended to NS binaries~\cite{hansen2001,lyutikov2011electro,lai2012dc,piro2012}, and also generalized for BHs, relying on ideas from the membrane paradigm \cite{thorne} to build a circuit model in black hole-neutron star (BHNS) binaries \cite{mcwilliams2011, lyutikov2011electro, lai2012dc, d2013big}. 

For a BHNS binary, one might associate the moving magnetic dipole (MD) effect with orbital energy being extracted from the NS, 
while the unipolar induction (UI) mechanism would operate to remove energy from the BH as it moves across the magnetic field of the NS.
In a neutron star-neutron star (NSNS) system, on the other hand, both mechanisms are expected to operate together at each NS, with intensities depending mainly on their relative magnetizations. That is, each star could produce the MD effect due to the orbital motion of its own magnetic moment, and could operate as UI as it moves across the magnetic field of its companion. 
Of course this is only schematic and non-trivial superposition of these two mechanisms, along with other possible effects (like, e.g., magnetospheric flares \cite{most2020}), would complicate the picture. 
The relative orientations of the magnetic moments (with respect to the orbital plane, and among them in the case of NSNS) and the role played by the spin of each compact object, complicates things even further.

Simulations of compact NS binaries in full general relativity (GR), paying attention to precursor EM signals, has been carried out (see, e.g.,~\cite{paschalidis2017} for a review).
Late-time inspiral phases of NSNS were considered in \cite{palenzuela2013electromagnetic, palenzuela2013linking, ponce2014} using general relativistic force-free (GRFF) simulations, broadly matching ideal magnetohydrodynamics (MHD) stellar interiors with an exterior force-free magnetosphere. Also, BHNS binary systems at fixed orbital separation --with and without BH spin-- were studied in \cite{paschalidis2013}, relying on a similar GRFF numerical approach. 
Overall, the EM luminosity found was consistent with estimations from the UI model\footnote{This holds only prior to the last few orbits before the merger in \cite{palenzuela2013electromagnetic, palenzuela2013linking, ponce2014}, in which the dynamics becomes more violent and nonlinear.} and the Poynting flux distributions were analyzed in both scenarios. 
Even though the MD mechanism has been mentioned in these articles as contributing into the total computed luminosity, typically less attention has been devoted to this effect in the literature as compared to the UI mechanism.
For NSNS binaries, further magnetospheric properties such as the formation of current sheets (CS) were reported \cite{palenzuela2013electromagnetic, palenzuela2013linking, ponce2014}, considering different relative magnetic strengths and orientations. 
However, in these previous studies, the details of the magnetosphere and the EM signals associated with its structure have not been considered in great depth.

One of the most challenging aspects of the problem is how --and how much of-- the available energy transferred to the plasma produces the emissions on the different EM bands.
Even for pulsars, for which the attempts to understand the principal magnetospheric properties has been done for many decades (e.g.,~\cite{contopoulos1999, mckinney2006relativistic, timokhin2006force, spitkovsky2006}), an ultimate answer to this question still remains elusive, although, of course, a lot can be learned from the development of pulsar theory in this respect (e.g.,~\cite{bai2010, uzdensky2013}). 
In recent years, there has been significant progress on particle-in-cell (PIC) simulations, which self-consistently model the regions of plasma production and particle acceleration (see, e.g.,~\cite{kalapotharakos2018, philippov2018, philippov2019pulsar}).

In this paper, we aim to further clarify the properties of the magnetosphere around an NS in a compact binary system.
We consider a single NS with dipolar magnetic fields and surrounded by a force-free plasma, in orbital motion. 
Our NS follows a trajectory, in flat spacetime, which mimics that of an NS about the center-of-mass (CoM) on a particular binary system. 
This way, we pay our attention to the MD energy extraction process and how this EM energy gets distributed within the magnetosphere; in a sense, decoupling it from the UI mechanism (or other curvature effects) involved in the binary. Such simplified setting allow us to conduct rather inexpensive, very accurate, numerical simulations for a detailed study of these systems. In particular, we investigate their magnetospheric features in close analogy to those of pulsars. And then, we use recent results of pulsar theory to infer possible EM signals from our numerical results. 
We see the expected $\gamma$ and $X$ rays luminosities estimated from the orbiting NS are rather weak, rendering their possible detections by current and near future facilities quite unlikely.
And thus, as observed in \cite{lyutikov2018electro}, the best chance to detect EM precursor signals from the inspiral phase of compact binary systems is from magnetospheric pulsar-like configurations producing coherent radio emission. 

The code used here to evolve the equations of force-free electrodynamics was first described in~\cite{FFE2}; 
and later extended in~\cite{NS}, where a careful treatment to handle the boundary conditions on the NS surface was presented, in contrast to the matching procedure used in previous GRFF simulations~\footnote{It is not clear whether such matching employed in the GRFF simulations can accurately represent the (approximately) perfectly conducting NS surface, which is the key condition for describing the magnetosphere around the NS.}. Since then, our code has been further tested and employed in other astrophysical scenarios \cite{Boost,Magnetar}, as well.

The paper is organized as follows. In Sec.~II we setup the problem and describe our numerical implementation. The results are presented in Sec.~III, first focusing on the magnetospheric properties of an NS in circular orbits, and then following inspiral trajectories associated with BHNS and NSNS binaries. Then, possible observational implications of our results are discussed. We summarize and conclude in Sec.~IV. Throughout this paper, $G$ and $c$ denote the gravitational constant and the speed of light, respectively.

\section{Setup}

\subsection{General Setting}

\begin{figure}[t]
\centering{
\includegraphics[scale=0.7]{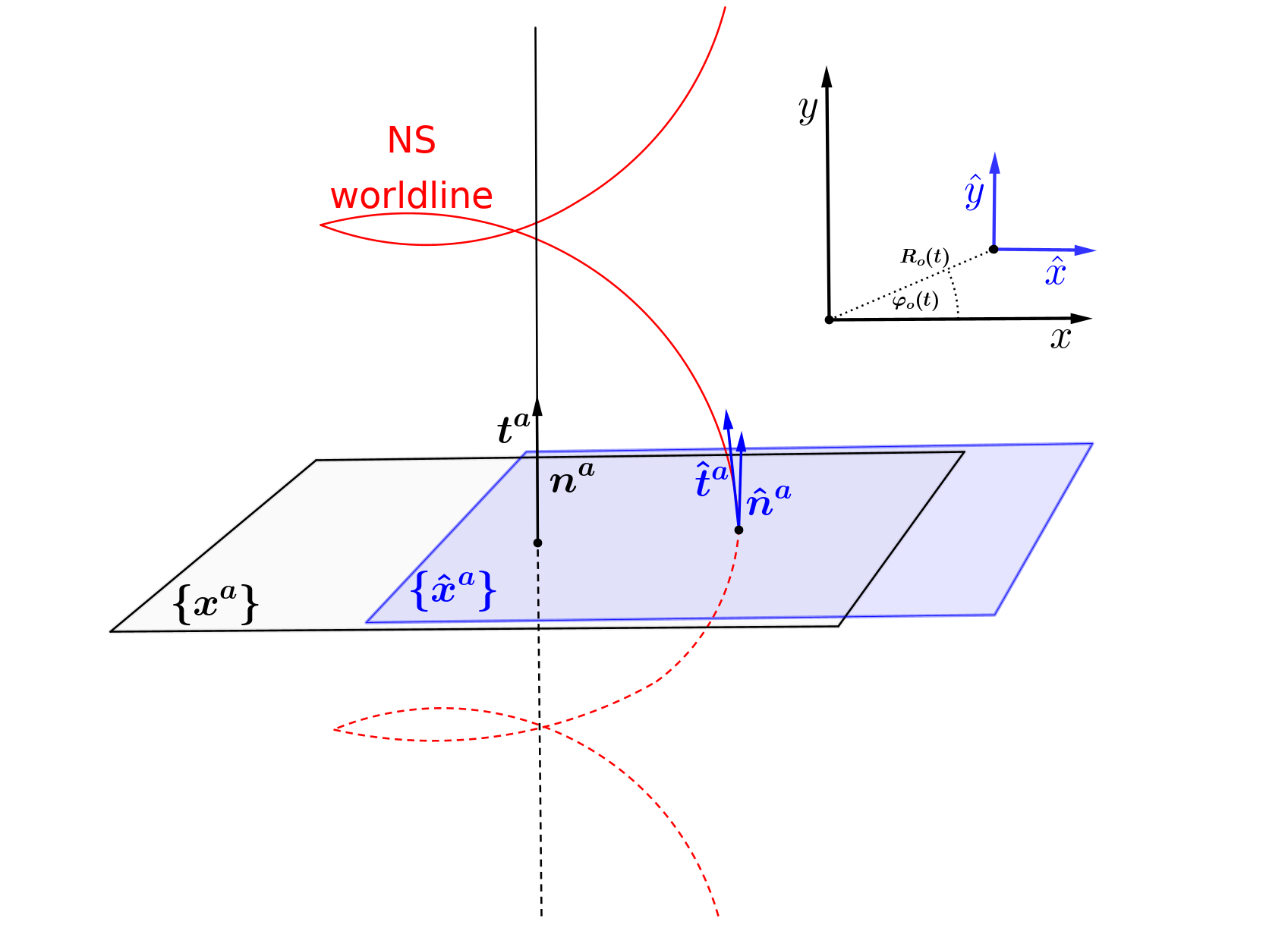}
\caption{\textit{Foliations and coordinates.} The two set of coordinates are illustrated, together with the worldlines of the NS (in red) and CoM (in black).}
\label{fig:frames}}
\end{figure}

The purpose of this paper is to clarify the magnetosphere around an NS in a compact binary system. Here we suppose that only the NS has a strong magnetic field and the magnetic field of its companion is much weaker. This is trivial for the BHNS case and would be a good approximation for the NSNS case because the first born NS in NSNS systems is likely to be weakly magnetized \cite{tauris2017}.  
We shall also assume that the spacetime is flat and that the NS moves on a given trajectory centered around the origin of a Cartesian coordinate system $x^a = \{t,x,y,z\}$, which would represent the CoM of the binary system.
The line element in this coordinates reads,
\begin{equation}\label{eq:flat}
 ds^2 = -dt^2 + dx^2 + dy^2 + dz^2, 
\end{equation}
thus, with $\alpha=1$, $\beta^i =0$ and $\gamma_{ij}=\delta_{ij}$ representing the lapse, shift and spatial metric, respectively. The trajectory is defined here by the radial distance $R_o (t)$ and the phase $\varphi_o (t)$ (being $\Omega_o (t) \equiv \dot{\varphi_o} (t) $ the associated angular velocity).
However, our numerical domain will be centered on the NS instead, thus describing the dynamics from an adapted foliation with coordinates $\hat{x}^a = \{\hat{t},\hat{x},\hat{y},\hat{z}\}$ (see Fig.~\ref{fig:frames}).
%
%
The coordinates transformation into this ``co-moving" foliation is defined by:
\begin{eqnarray}
 && t = \hat{t}, \quad z = \hat{z}, \nonumber \\
 && x = \hat{x} + R_{o}(\hat{t}) \cos (\varphi_o (\hat{t})), \\
 && y = \hat{y} + R_{o}(\hat{t}) \sin (\varphi_o (\hat{t})),  \nonumber
\end{eqnarray}
and thus,
\begin{eqnarray*}
 && dt = d\hat{t}, \quad dz = d\hat{z},  \\
 && dx = d\hat{x} + \left[ \dot{R}_{o}(\hat{t}) \cos (\varphi_o (\hat{t})) - R_{o}(\hat{t}) \Omega_{o}(\hat{t}) \sin (\varphi_o (\hat{t}) ) \right]  d\hat{t},  \\
 && dy = d\hat{y} + \left[ \dot{R}_{o}(\hat{t}) \sin (\varphi_o (\hat{t})) + R_{o}(\hat{t}) \Omega_{o}(\hat{t}) \cos (\varphi_o (\hat{t}) ) \right]  d\hat{t},  
\end{eqnarray*}
Therefore, the line element \eqref{eq:flat} in the new coordinates is,
\begin{equation}\label{eq:metric}
 d\hat{s}^2 = \left( \hat{\beta}^2 - \hat{\alpha}^2 \right) d\hat{t}^2 + 2 \hat{\beta}_i d\hat{x}^i d\hat{t} + \hat{\gamma}_{ij} d\hat{x}^i \hat{x}^j,
\end{equation}
where again $\hat{\alpha}=1$ and $\hat{\gamma}_{ij} = \delta_{ij} $. However, there is now a non-vanishing shift vector accounting for the difference among the NS worldline tangent $\hat{t}^a$, and the hypersurface normal $\hat{n}^a$ in this foliation. That is,
\begin{equation*}
 \hat{\beta}^i =  \left\lbrace \dot{R}_o  \cos \varphi_o  - v_o \sin \varphi_o , \dot{R}_o  \sin \varphi_o + v_o \cos \varphi_o, 0 \right\rbrace,
\end{equation*}
where we have dropped all time dependencies and defined $v_o := R_o \Omega_o$ to simplify the notation. We will consider essentially two types of trajectories in this work: (i) a purely circular motion at a constant radius $R_o$ and orbital frequency $\Omega_o$; and (ii) quasi-circular inspiral orbits which mimic those of an NS in a binary system close to merger.

The boundary condition at the stellar surface is derived by assuming the perfectly conducting condition,
\begin{equation}
 0 = F_{ab} \hat{t}^a = F_{ab} ( \hat{\alpha} \hat{n}^a + \hat{\beta}^a ),
\end{equation}
which can be easily generalized to incorporate the NS spin at frequency $\Omega_*$ by,
\begin{equation}
 F_{ab} ( \hat{t}^a + \Omega_* \hat{\phi}^a ) = 0,
\end{equation}
where $\hat{\phi}^a \equiv (\partial_{\hat{\phi}})^a $. Thus, the resulting condition on the electric field measured by a fiducial observer in this adapted foliation (i.e. $\hat{E}_a := F_{ab} \hat{n}^a$), can be written:
\begin{equation}\label{eq:BC}
 \hat{E}^i =  \epsilon^{i}_{\phantom{i}jk} (\hat{\beta}^j + \Omega_* \hat{\phi}^j ) \hat{B}^k .
\end{equation}

The initial configuration is taken to be a magnetic dipolar field (of dipole-moment $\mu$) and vanishing electric field. The NS is gradually set in motion (at fixed $R_{o,0}$) until it reaches the desired initial orbital frequency $\Omega_{o,0}$ after some time $t=t_0$; from then on, the system follows its prescribed trajectory.
We focus primarily on the cases where the magnetic axis is aligned with the orbital angular momentum, but other scenarios in which these two axis are not aligned are considered as well. In our setup, the $z$ axis is always perpendicular to the orbital plane, so the misalignment is attained by just tilting the magnetic moment by an angle $\chi$ along the $x$-$z$ plane.

\subsection{Numerical Implementation}

We evolve a particular version of force-free electrodynamics derived in \cite{FFE}, which has some improved properties in terms of well posedness and involves the full force-free current density.
More concretely, we shall consider the evolution system given by Eqs.~(8)--(10) in \cite{NS}.
Our numerical scheme to solve these equations is based on the \textit{multi-block approach} \cite{Leco_1, Carpenter1994, Carpenter1999, Carpenter2001}, in which the numerical domain is built from several non-overlapping grids where only grid-points at their boundaries are sheared. The equations are discretized at each individual subdomain by using difference operators constructed to satisfy summation by parts. In particular, we employ difference operators which are eighth-order accurate on the interior and fourth-order at the boundaries. 
Numerical dissipation is incorporated through the use of adapted Kreiss-Oliger operators. These compatible difference and dissipation operators were both taken from \cite{Tiglio2007}. A fourth order Runge-Kutta method is used for time integration. 

We solve the force-free equations in a region between an interior sphere at radius $\hat{r}=R_*$ that represents the NS surface (i.e., $R_*$ denotes the NS radius), and an exterior spherical surface located at $\hat{r}\sim 75 R_*$.
The domain is covered by a total of $6 \times 12$ subdomains, with $6$ patches to cover for the angular directions and $12$  being the number of spherical shells expanding in radius. These spherical shells do not cover regions of identical radial extension, having more resolution near the inner boundary than in the asymptotic region: from layer to layer, the radial resolution is decreased by a factor $1.3$.
Typically we adopt a resolution with total grid numbers of $N_{\hat{\theta}} \times N_{\hat{\phi}} \times N_{\hat{r}}$ with $N_{\hat{\phi}} = 2 N_{\hat{\theta}} = 240$, while $N_{\hat{r}}$ is taken so as to satisfy $\Delta \hat{r} \lesssim 0.7 \, \hat{r}  \Delta \hat{\theta}$ everywhere in the domain. 
Here $\Delta \hat{r}$ and $\Delta \hat{\theta}$ denote the grid spacing for $\hat r$ and $\hat \theta$, respectively.

As already mentioned, the stellar surface is assumed to behave as an idealized perfect conductor. 
Thus, the normal component of the magnetic field is set to its dipole value assumed from the stellar interior and the electric field is prescribed according to Eq.~\eqref{eq:BC} in order to represent its orbital motion and spin.
The electric field components are imposed, by means of the \textit{penalty method} \cite{Carpenter1994, Carpenter1999, Carpenter2001}, fixing the incoming physical modes to a particular combination of outgoing modes.  
At the outer boundary, on the other hand, we set maximally dissipative (no-incoming) conditions to allow all perturbations to propagate away. 
The numerical implementation of such boundary conditions has been detailed in \cite{NS} (in particular, Sec.~II-C and Appendix), so we recommend the interest readers to refer there for further details.
In order to handle CS, for which the force-free approximation breaks down, we use a standard approach in which the electric field is effectively dissipated to maintain the condition that the plasma is magnetically dominated (i.e., $B^2 -E^2 >0$), as discussed in \cite{FFE2} (see also \cite{komissarov2004}).

\subsection{Analysis Quantities}

We would like to monitor the EM energy and its associated fluxes.  
In force-free electrodynamics the four-momentum, $p^a = -T_{EM}^{ab} t_b$, is conserved (i.e. $\nabla_a p^a = 0$)~\footnote{This equation is satisfied except for CS, where dissipation occurs.} 
in the stationary spacetime. 
In the co-moving coordinates $\{ \hat{x}^a \}$, it reads:
\begin{eqnarray}
 p^{a'} &=& - T_{EM}^{a'b'} t_{b'} = - T_{EM}^{a'b'} \hat{n}_{b'} \nonumber\\
	&=& \frac{1}{2} ( \hat{E}^2 + \hat{B}^2 ) \, \hat{n}^{a'} - \hat{S}^{a'} .
\end{eqnarray}
Hence, we measure:
\begin{equation*}
 E(\hat{t}) := \int_{\Sigma_{\hat{t}}} \hat{\mathcal{E}} \sqrt{\hat{\gamma}}\, d^{3}\hat{x} \text{,  }\quad L (\hat{t}, \hat{r}) := \oint_{\hat{r}} \hat{\mathcal{F}}_E \, \sqrt{-\hat{g}}  \, d^{2}\hat{x}, 
\end{equation*}
where the Poynting luminosity $L$ is integrated on spherical surfaces of radius $\hat{r}$ around the NS and,
\begin{eqnarray}
\hat{\mathcal{E}} &:=& -p^{a'} \hat{n}_{a'} = \frac{1}{2} ( \hat{E}^2 + \hat{B}^2 ),  \\
\hat{\mathcal{F}}_E  &:=& p^{a'} (d\hat{r})_{a'} = - \frac{1}{2} ( \hat{E}^2 + \hat{B}^2 ) \, \hat{\beta}^{\hat{r}} - \hat{S}^{\hat{r}} ,
\end{eqnarray}
with $\hat{S}^{i} := \epsilon^{ijk} \hat{E}_j \hat{B}_k$ being the spatial Poynting vector.

We are also interested in monitoring the charge distribution and electric currents present during the dynamics. 
Thus, we shall look at the force-free current density along the magnetic field, as seen by a fiducial observer $\hat{n}^{a}$,
\begin{equation}
 \hat{j}_{\parallel} = (\hat{B}_k \hat{\beta}^k)  \hat{\rho}_c +  \hat{B}_k \mathcal{D}_j H^{kj} + \hat{E}_k \mathcal{D}_j G^{kj} ,
\end{equation}
where we denoted $\mathcal{D}_{j}(\cdot):=\frac{1}{\sqrt{\hat{\gamma}}} \partial_{j}( \sqrt{\hat{\gamma}} ~ \cdot ~)$ 
and,
\begin{eqnarray}
H^{ij} &:=&  \hat{E}^{i} \hat{\beta}^{j} - \hat{E}^{j} \hat{\beta}^{i} + \epsilon^{ijk} \hat{B}_{k}, \\
G^{ij} &:=&  \hat{B}^{i} \hat{\beta}^{j} - \hat{B}^{j} \hat{\beta}^{i} - \epsilon^{ijk} \hat{E}_{k},
\end{eqnarray}
with $ \hat{\rho} = \mathcal{D}_j \hat{E}^j $ being the charge seen by this observer.

Actually, it is worth mentioning at this point that since in this case $n^{a}$ and $\hat{n}^{a}$ represent exactly the same vector field, the splitting of the electromagnetic tensor in its magnetic and electric components is the same. Hence, the switch from one description to the other in terms of the electric and magnetic fields is quite direct. Since the vector transformation for the spatial index is trivial, one is only left with the appropriate displacement of the point where the field is evaluated. Therefore, although we employ the NS frame $ \{ \hat{x}^a\}$ to evolve the fields, we use the ``CoM'' coordinates $ \{ x^a\}$ to plot all the relevant quantities and describe our results.

We also consider the Lorentz invariant quantity,
\begin{equation}\label{eq:currents}
 \varrho := \pm \sqrt{|\rho^2 c^2 - j^2 |}, 
\end{equation}
where the sign is chosen plus for timelike and minus for spacelike currents, as in \cite{bai2010}.
Counter-streaming of different signs of charge is required at regions having negative values of this quantity, which may lead to plasma instabilities and dissipation (see, e.g.,~\cite{lyubarskii1996, gruzinov2008, bai2010}).


\section{Results}

Our interest in this work is centered on the last few orbits of an NS on a compact binary system, until the orbit reaches an innermost stable circular orbit or the NS gets tidally disrupted.    
Thus, we start by a detailed study of the magnetospheric properties of the NS in a circular orbit, choosing parameters in a relevant range: i.e., $R_o \sim (2$--$6) R_* $ and $ v_o := R_o \Omega_o =(0.1$--$0.4) \, c$.
We first focus on the case in which the magnetic moment is aligned to the orbital angular momentum (i.e., $\chi=0$) and the NS is not spinning (i.e., $\Omega_* = 0 $). 
Our simulations always relax to stationary states presenting similar features to those of pulsar magnetospheres, with strong equatorial CS.
We analyze the electric charge/current distributions of the surrounding plasma, along with the resulting Poynting flux luminosity  inside the orbital light cylinder.
A comparison with vacuum magnetospheres, within the same setting, is also included here.
Then, both NS spin and the misalignment of the magnetic axis are incorporated into the picture. 
We vary these parameters (i.e., $\Omega_*$ and $\chi$) independently, and observe their impact on the solutions. 
Later, we consider representative quasi-circular inspiral orbits, taking relevant values for the parameters in the contexts of BHNS and NSNS binaries, and connecting with our previous luminosity estimations.
Finally, we elaborate on the implication of our results to EM observations.

\begin{figure*}[!ht]
\centering{
\includegraphics[scale=0.22]{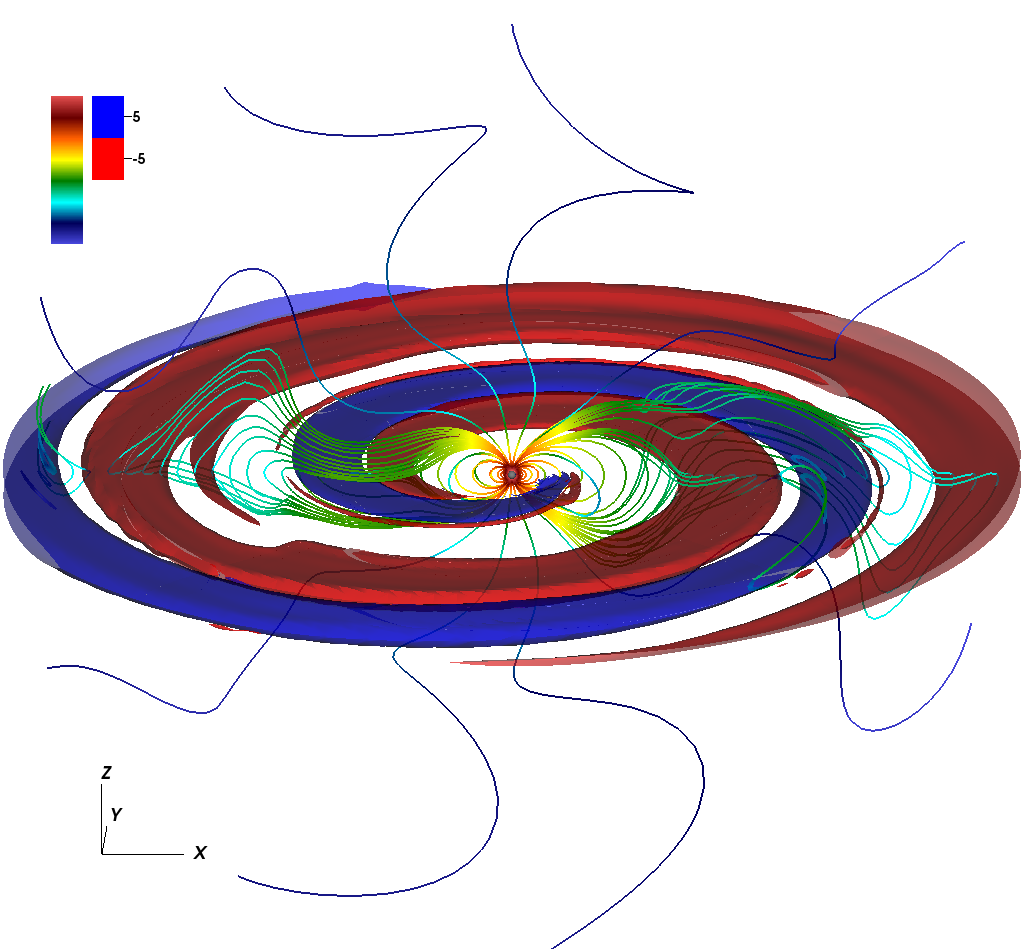}
\includegraphics[scale=0.22]{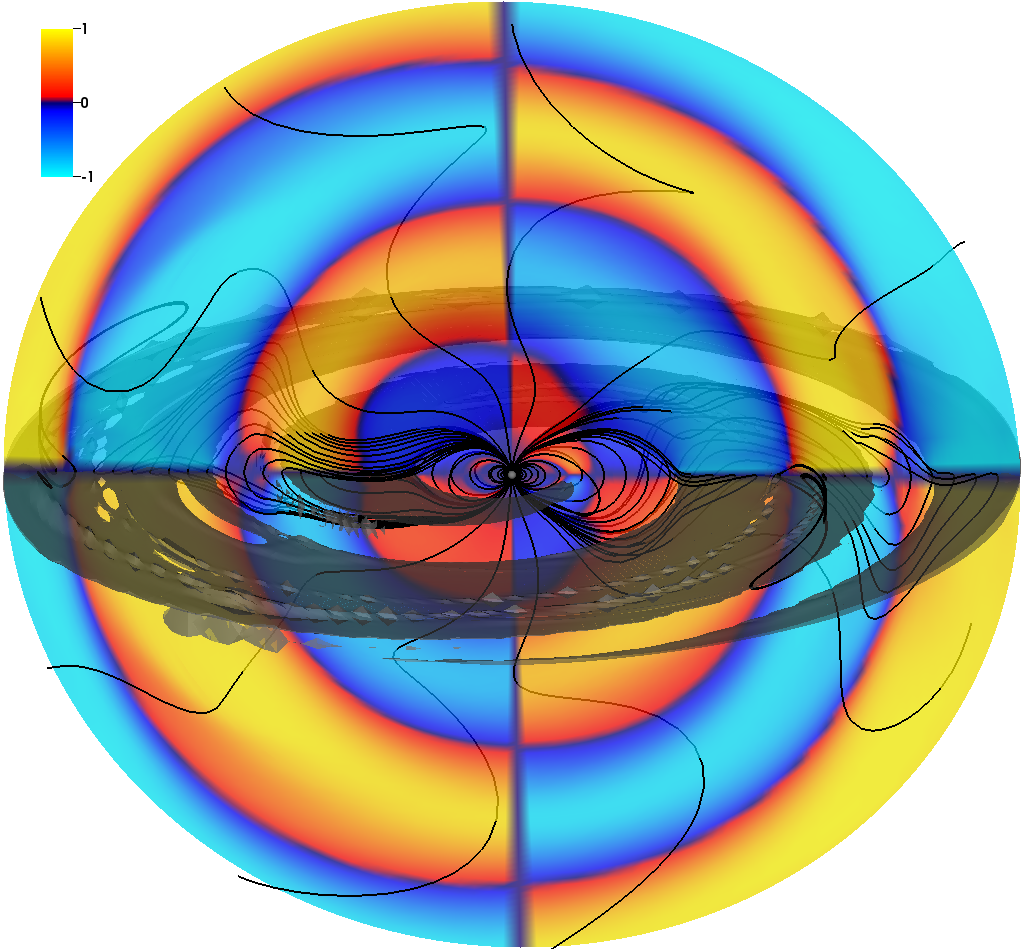}
\caption{\textit{Magnetosphere for a circular orbit with $R_o \Omega_o = 0.25 c$ and $R_o = 2.5 R_*$ ($\chi=0$, $\Omega_* =0$) after 3 periods.}
Left panel: representative magnetic field lines (with the colors indicating its magnitude in logarithmic scale); 
along with two contours of the electric charge density, normalized by the Goldreich-Julian value $\Omega_o B/2\pi c$. 
Right panel: normalized toroidal component of the magnetic field, $B^{\phi}/B$, at the $y=0$ plane;
together with a contour of $B^2 - E^2 \sim 0.04 \, B^2$ (in gray), signaling the presence of an spiral CS over the orbital plane.
Same representative magnetic field lines are also indicated in the figure (in solid black). 
}
 \label{fig:CS}}
\end{figure*}

\subsection{Circular Orbits with no spin}

First, we consider the case in which the NS is in a circular orbit with no spin. In numerical computation, the NS is set into circular orbital motion by gradually bringing the angular frequency to its final value $\Omega_o$.
The motion generates disturbances on the EM field that propagates into the surrounding plasma; when these waves return to the NS surface, they are reflected due to the perfectly conducting boundary condition. After an initial transient of about 2 orbits, these perturbations --continuously injected from (and reflected at) the stellar surface-- equilibrate within the magnetosphere and lead to a quasi-stationary solution.
Such configuration is illustrated by the 3D plots in Fig.~\ref{fig:CS}, where several representative magnetic field lines are shown. 
The injected Alfven waves twist the magnetic field lines, as the non-spinning NS follows the circular orbit, producing a pattern of alternate signs on their toroidal components, like the one depicted in the right panel of Fig.~\ref{fig:CS}.
Waves launched from magnetic footprints in opposite hemispheres of the star meet at the dipole equator, inducing sharp discontinuities on the toroidal field across the orbital plane.  
These discontinuities form an spiral CS, which is represented by the contour plots in the figures. 
Electric field is being effectively dissipated at these regions, in order to locally maintain a state in which $B^2 - E^2 \gtrsim 0$.
This is a rather standard strategy employed in force-free electrodynamics simulations to avoid violations of the magnetic dominance condition $B^2 - E^2 > 0$ (see \cite{komissarov2004} for a physical justification).
Thus, a small value of this Lorentz invariant quantity is a convenient indicator to illustrate CS in this context (right panel). 
On the other hand, left panel of Fig.~\ref{fig:CS} shows that intense charge density (exceeding the Goldreich-Julian value) develops at the CS, possessing two main components of opposite sign over the spiral arms. As can be noticed, magnetic reconnection take place near the transition from negative to positive charge, producing and ejecting closed magnetic loops outwards.

The spiral structure of the magnetic field, clearly manifest on the equatorial CS, is also found for other relevant quantities such as electric charge/currents and Poynting flux density distributions; 
thus, reflecting the helical symmetry of the problem in this particular setting. The EM solution looks essentially static from a co-rotating frame.

As we shall see in more detail later, these configurations resemble in many aspects those of pulsar magnetospheres, 
with CS outside the wave zone $\sim c/\Omega_{o}$, and energy being extracted (in this case, kinetic energy from the orbital motion) and carried away by the surrounding plasma. 
\begin{figure}
\centering{
\includegraphics[scale=0.22]{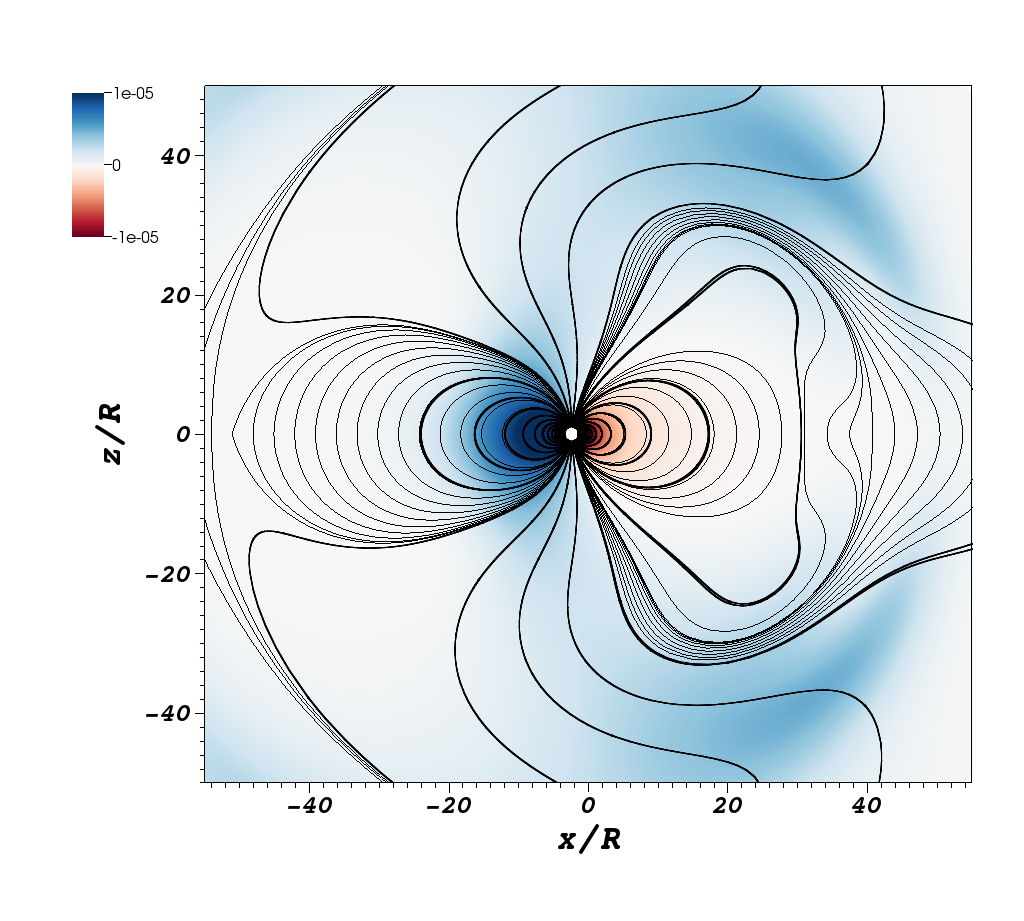}\\
\includegraphics[scale=0.22]{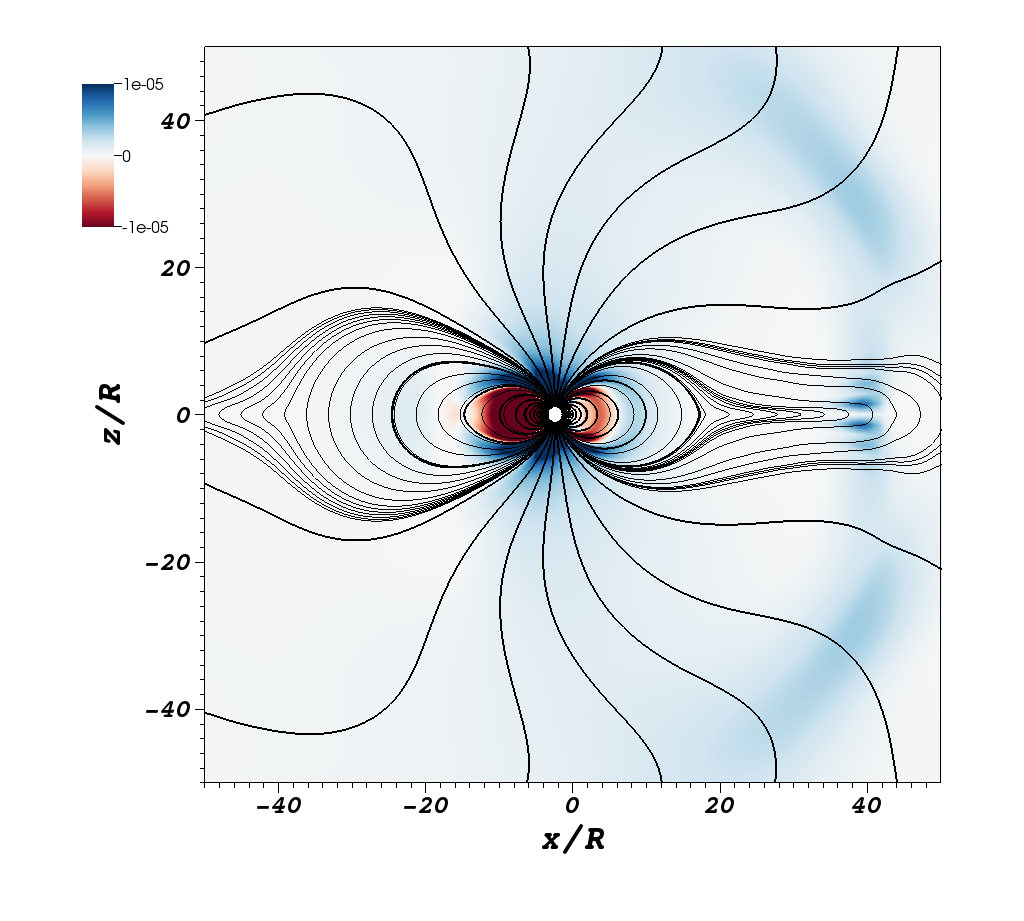}
\caption{\textit{Electromagnetic energy fluxes for a circular orbit with $R_o \Omega_o = 0.25 c$ and $R_o = 2.5 R_*$ ($\chi=0$, $\Omega_* =0$).} Radial Poynting flux distribution (color scale) and magnetic field topology (solid lines) after $2.5$ periods, for vacuum (top panel) and force-free (bottom panel) magnetospheres are shown on the $y=0$ plane. The NS is located at $x=-R_o$ and $z=0$. (Note that $R$ in the figures denotes $R_*$). 
}
 \label{fig:comp_PF}}
\end{figure}
Figure \ref{fig:comp_PF} shows the topology of the magnetic field lines after $2.5$ orbits on the $y=0$ plane (i.e., the one containing both the NS and CoM). 
Vacuum and force-free solutions are included for a comparison. We remark that even though the driving mechanism is the same (i.e., the effect of a moving MD moment) and both system produce comparable radial Poynting flux intensities (shown in color scale in the figure), the presence of the plasma changes the topology of the magnetic field and the form in which the EM energy is distributed and transported. In particular, the force-free plasma is essential ingredient for CS formation.

\begin{figure}
\centering{
\includegraphics[scale=0.6]{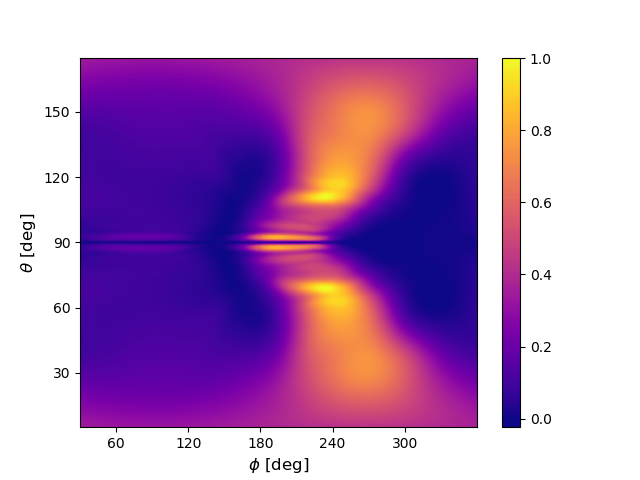}
\caption{\textit{Angular distribution of the electromagnetic flux for a circular orbit with $R_o \Omega_o = 0.25 c$ and $R_o = 2.5 R_*$ ($\chi=0$, $\Omega_* =0$).}
The radial Poynting flux density after $2.5$ orbits is plotted on a spherical surface at radius $\hat{r}\sim 60 R_*$ (i.e., in the wave zone).
The flux is normalized with its peak value.}
 \label{fig:PF_60R}}
\end{figure}

We further analyze the Poynting flux distribution of the force-free solutions over spherical surfaces located in the outer wave zone.
One such EM flux density, after $2.5$ orbits at a sphere of radius $\hat{r} \sim 60 R_*$, is shown in Fig.~\ref{fig:PF_60R}.
As can be seen from the plot, the flux is concentrated along a broad beam of $\sim 60$\textdegree$-75$\textdegree \, in the azimuthal direction and within $\sim 60$\textdegree \, from the orbital plane.
There is a visible feature in the distribution at $\hat{\theta} \sim 90$\textdegree, reflecting the magnetic reconnections occurring at the equatorial CS.
Once the magnetosphere has settled, these structures in the Poynting flux become stationary, just co-rotating with the orbit. And thus, the flow of EM energy produces a lighthouse effect at orbital frequency.
Besides the specific details of the distribution presented here, our results are in good qualitative agreement with the ones obtained from GRFF simulations of binary systems involving a non-spinning BH companion \cite{paschalidis2013},
and also for a weakly magnetized NS companion in \cite{palenzuela2013electromagnetic, palenzuela2013linking}.

\begin{figure}
\centering{
 \includegraphics[scale=0.33]{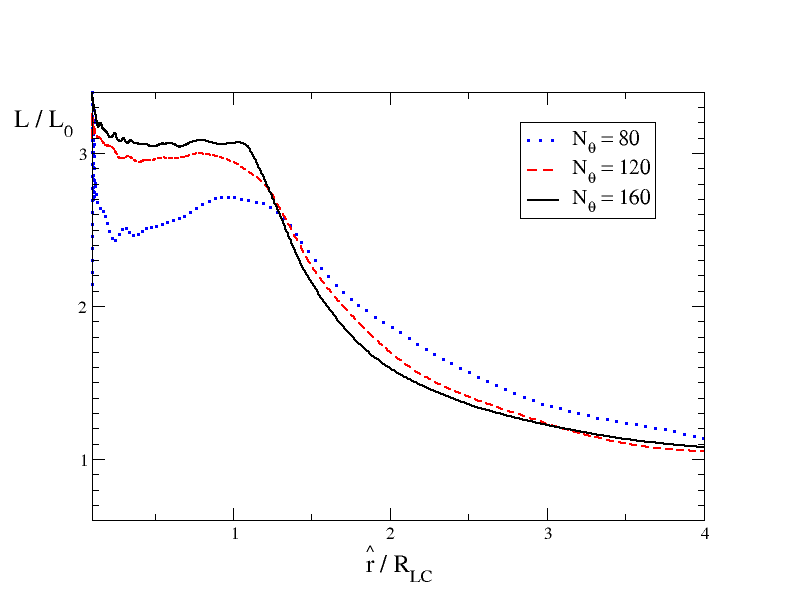}

\caption{\textit{Convergence of the luminosity}. Normalized luminosity, for a circular orbit, as a function of integration radius for three different numerical resolutions, $N_{\theta} = \{ 80, 120, 160 \}$. } 
\label{fig:L_conv}
}
\end{figure}

We integrate the luminosity for the late time solutions at different radius and normalize them with the EM luminosity by the 
MD radiation formula (see, e.g.,~\cite{ioka2000}), 
\begin{eqnarray}
L_{0} =\frac{4}{15c^5} \mu^2 R_{o}^2 \Omega_{o}^6.  \label{eq:zeroth} 
\end{eqnarray}
A typical radial distribution of the luminosity can be seen in Fig.~\ref{fig:L_conv}, where we plot integrated Poynting flux through concentric spheres around the NS, as a function of radius. As in pulsars, the luminosity is constant up to the light cylinder $R_{LC} \equiv c / \Omega_o$, where dissipation at the CS begins. 
Different numerical resolutions were considered in order to test its convergence. Throughout this work, we have employed the intermediate resolution  $N_{\theta} = 120$ for all the simulations, which is practically converging (it differs in less than $4\%$ with respect to the higher resolution one). We note that such resolution required in the present orbital setting is larger than the one we typically needed for pulsars (see Fig.2 in \cite{NS}), where $N_{\theta} = 80$ was enough to resolve even for the misaligned configurations. 
Notice that dissipation taking place in the region $\hat{r} \sim (1$--$4) R_{LC}$ represents here $65\%$ of the luminosity. These are the typical percentages that we get for most of the circular orbits explored, while the values obtained for pulsars were instead closer to $40\%$ (within the same region)~\footnote{Here, we must emphasize that the amount of dissipation depends on the numerical prescription to deal with CS, where the force-free approximation breaks down. So, this values should only be taken as an indication of how strong these CS are, and not as quantitative astrophysical numbers. For the quantitative study of the amount of dissipation, a different approach like, e.g., PIC simulations would be needed.}. 
We have also tested our outer boundary conditions, by considering different radial locations for the outer surface of the computational domain. The solutions found at the overlapping regions are essentially identical, both qualitatively and quantitatively. 

\begin{figure}
\centering{
 \includegraphics[scale=0.33]{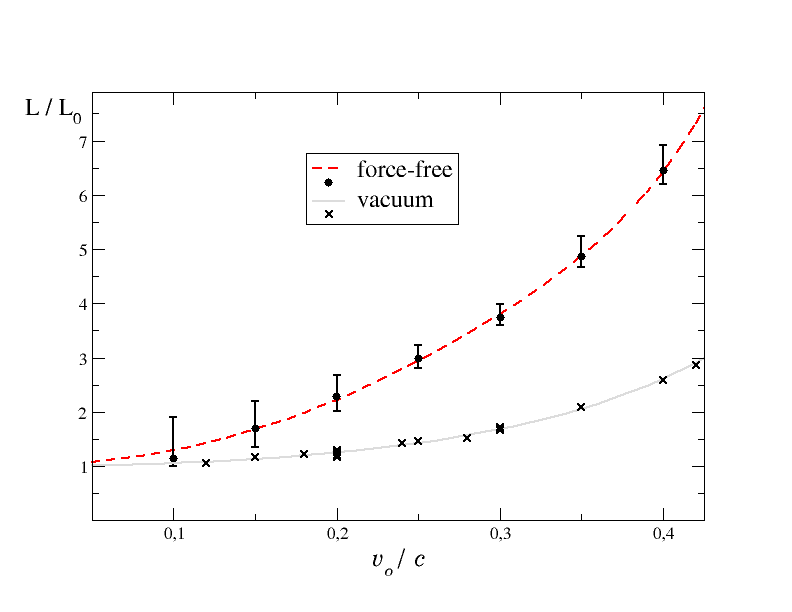}

\caption{\textit{Luminosity for circular orbits in force-free and vacuum cases}. The luminosity normalized by $L_0 := \frac{4}{15} \mu^2 R_{o}^2 \Omega_{o}^6$ as a function of $v_o := R_o \Omega_o$ is shown.
The black dots show the numerical force-free values, accompanied by a guideline (see Eq.~\eqref{eq:ffe}) in dashed red. The black crosses refer to numerical data from vacuum solutions, with a reference curve, $\gamma_{o}^{11}$, in grey ($\gamma_o$, being the Lorentz factor).} 
\label{fig:L_scaling}
}
\end{figure}

The scaling of the total luminosity as a function of the orbital velocity is analyzed and summarized in Fig.~\ref{fig:L_scaling}.
For the vacuum magnetosphere we find that $ L/L_0 \approx \gamma_{o}^{11}$ (where $\gamma_o \equiv \displaystyle\frac{1}{\sqrt{1- (v_o /c)^2}}$ is the Lorentz factor of the orbital velocity) approximates the numerical results quite well (perhaps ``accidentally'').
On the other hand, the following expression is used as a guideline for the luminosity in the force-free simulations:
\begin{eqnarray}\label{eq:ffe}
 \frac{L}{L_0} &\approx& 1 + 28 \left( \frac{v_o}{c}\right)^2 + 93 \left(\frac{v_o}{c}\right)^4 \nonumber\\
 &&  - 1135 \left(\frac{v_o}{c}\right)^6 + 4814 \left(\frac{v_o}{c}\right)^8 .
\end{eqnarray}
The error bars in the force-free data indicate certain dispersion found in the values of $L$ within the light cylinder. 
These errors become more significant at lower values of $v_o$ --as can be expected due to the enlargement of the wave zone--, for which a higher resolution (than the one employed here for all the runs) would be needed. 

We note that the luminosity does not depend on the other dimensionless parameter of the problem, i.e., $R_o / R_*$.
Within our numerical error, we could not find significant deviations on the ratio $L / L_0$, when taking different values of this parameter at fixed $v_o$.  

\begin{figure*}
	\centering
\includegraphics[scale=0.163]{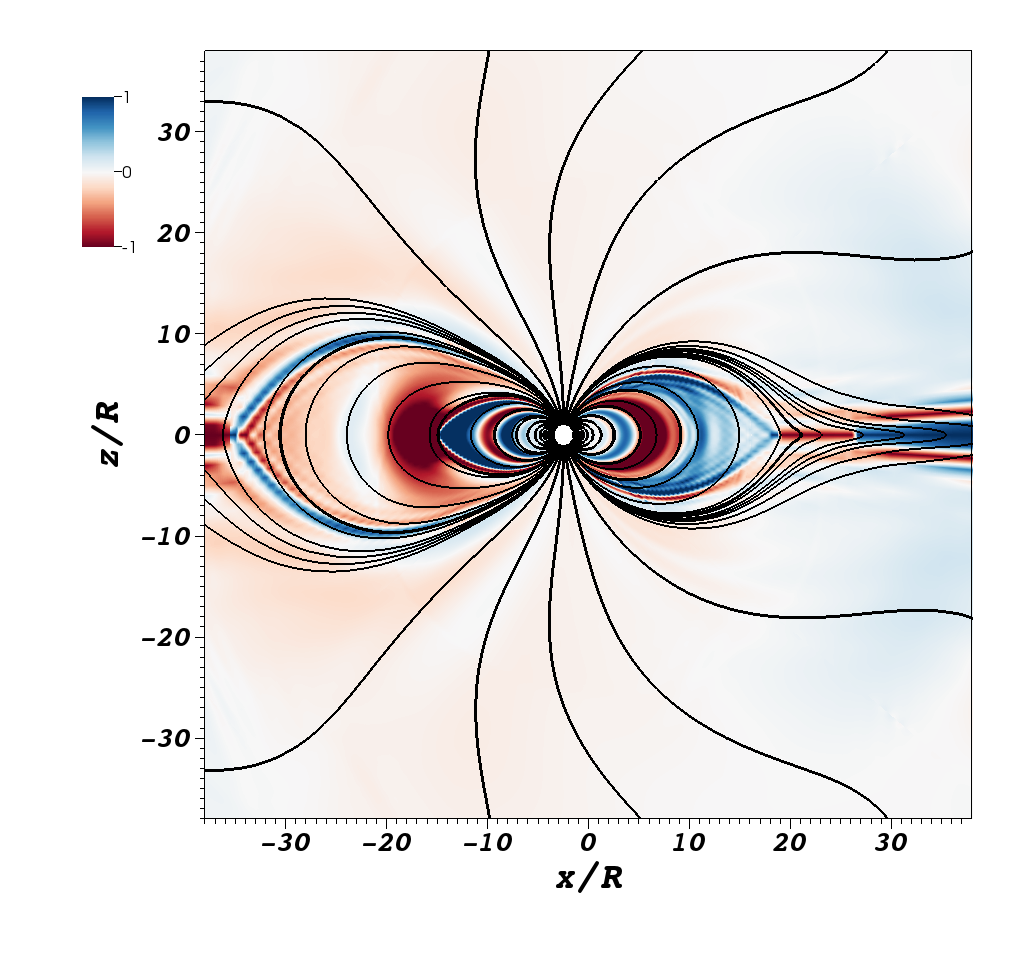}
\includegraphics[scale=0.163]{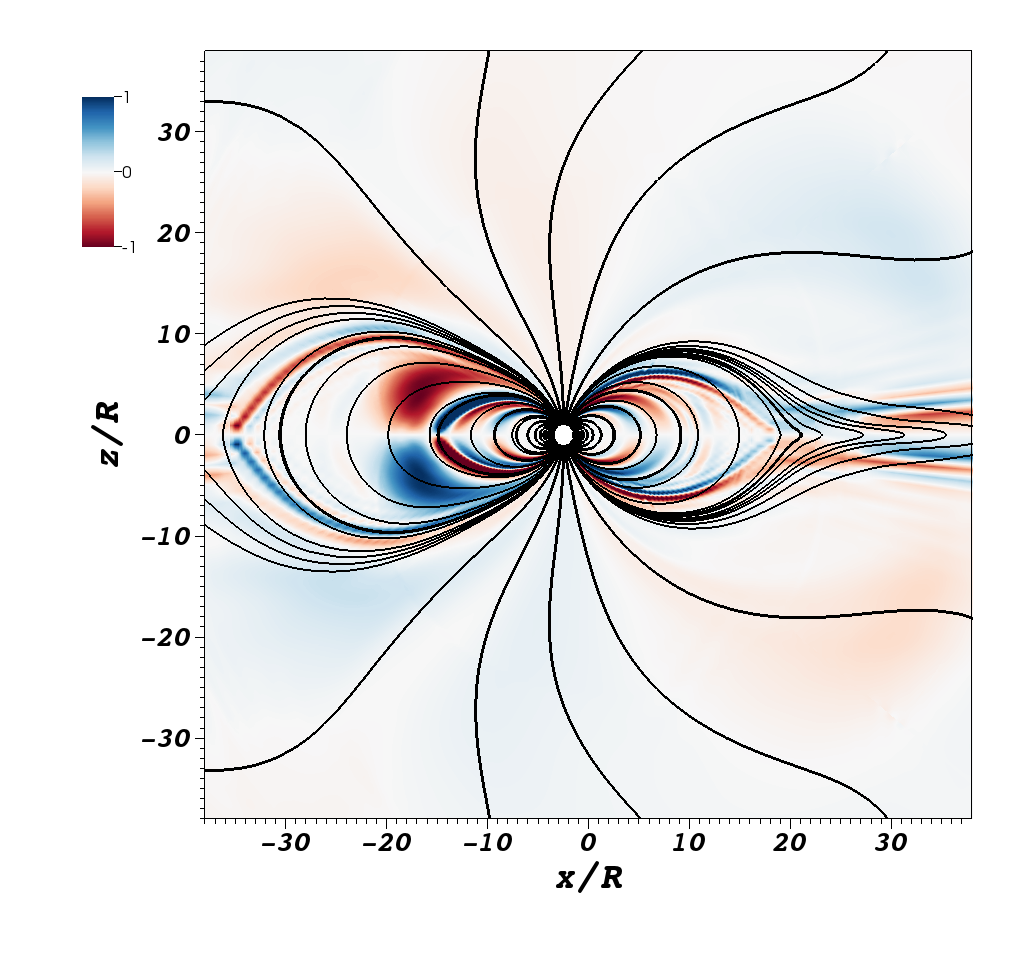}
\includegraphics[scale=0.163]{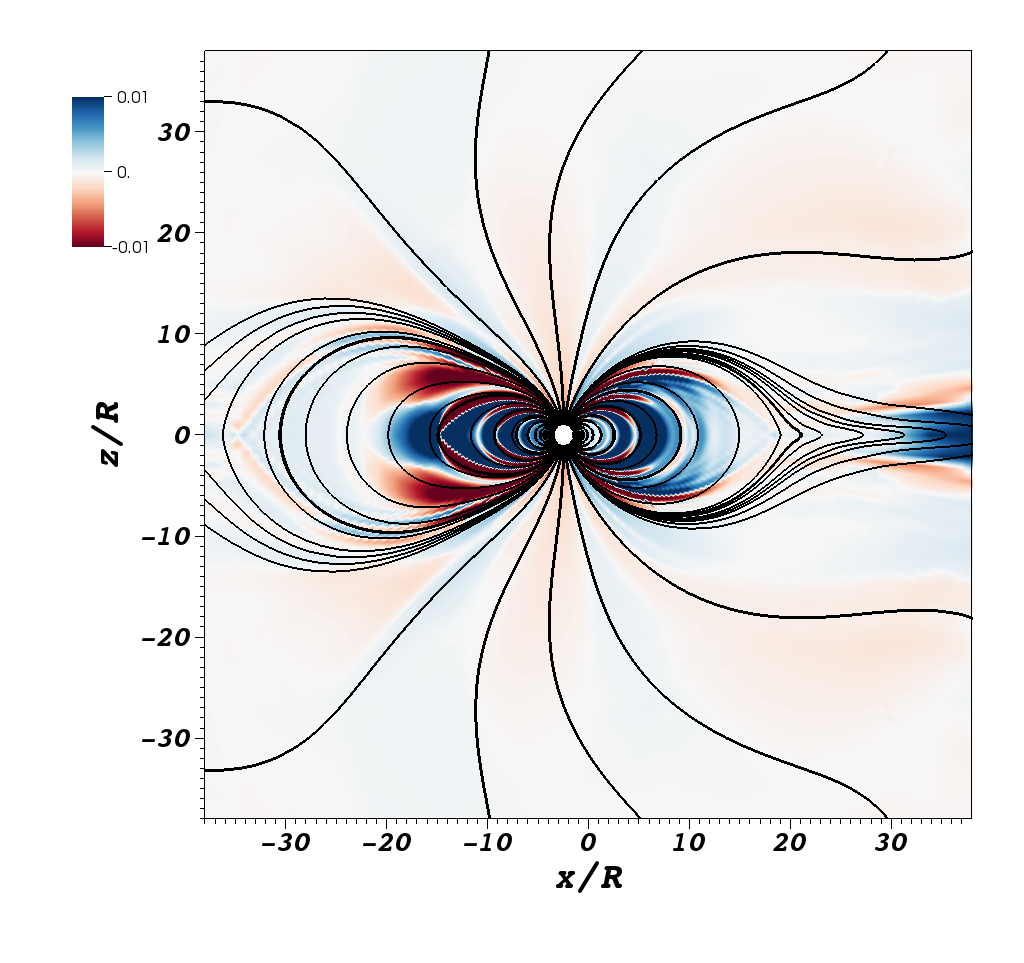}	
	\caption{\textit{Several quantities for a circular orbit with $R_o \Omega_o = 0.25 c$ and $R_o = 2.5 R_*$ ($\chi=0$, $\Omega_* =0$).} Numerical solution at $t=2.5$ periods on the $y=0$ plane. Black lines in all the plots describe the magnetic field component along the plane. Charge density distribution (left panel) and electric currents along magnetic lines (middle panel) are normalized by $\Omega_o B / 2\pi c$ and $\Omega_o B / 2\pi$, respectively. 
	Right panel: parameter $\varrho$ from Eq.~\eqref{eq:currents}  (multiplied by $r^2$ to improve the contrast), showing timelike (blue) and spacelike (red) currents. 
	}
	\label{fig:co-plane}
\end{figure*}

The magnetospheric structure after $2.5$ orbital periods is displayed in Fig.~\ref{fig:co-plane}, in which relevant aspects of the solution are represented on the $y=0$ plane (the plane containing both the NS and the CoM). 
There, electric charge and current distributions are displayed (left and middle panels); as well as the Lorentz invariant quantity $\varrho$ from Eq.~\eqref{eq:currents}. Magnetic field lines projected onto the plane are also shown in all these plots for a reference.

First, we notice that similarly to aligned pulsar solutions, there are strong current layers that form a Y-point with the equatorial CS (see middle panel). The magnetic field lines beyond this region are not necessarily open, since the equatorial CS do not extend continuously outside, but instead posses an spiral structure (as previously seen in Fig.~\ref{fig:CS}).
%
In contrast to a spinning NS, we find here that the orbital motion induces a pattern of alternating signs in the charges and parallel currents inside the light cylinder.
This is qualitatively consistent with the current distributions shown in Fig.~12 of~\cite{palenzuela2013linking}, especially for the case of a weakly magnetized NS companion ``$U/u$'' where we can interpret this alternate pattern as arising from the orbit (MD) effect and the strong currents connecting the two NSs as representing the UI mechanism.
Similar pattern is also found for the quantity $\varrho$ (right panel), meaning that the four-current alternates character from timelike to spacelike, the later being the regions where counter-streaming of charged particles would be expected.
Finally, and arguably the most relevant difference of the orbital case with respect to pulsar magnetosphere solutions 
is the fact that there is almost no charges present nor currents flowing along the magnetic field lines in the polar region of the NS.

\subsection{Spin effects}\label{spineff}

\begin{figure}
\centering{
\includegraphics[scale=0.33]{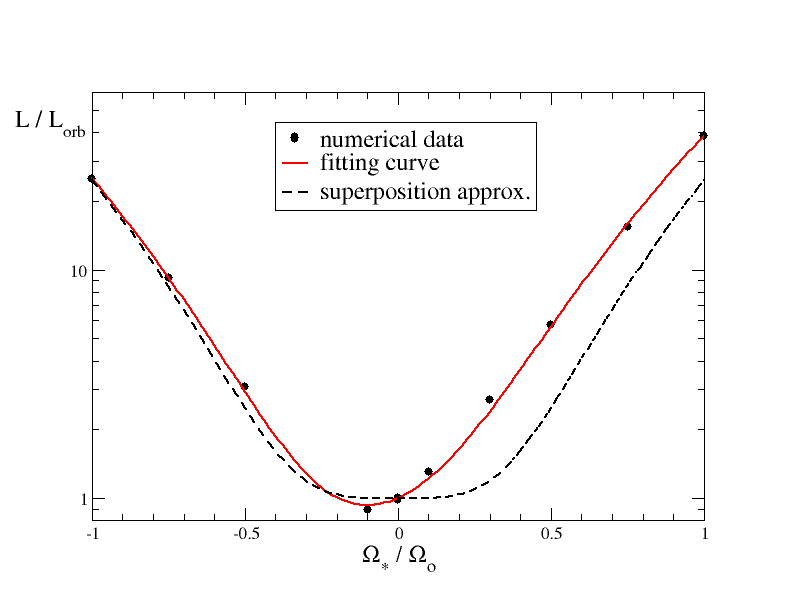}
\caption{\textit{Effects of neutron star spin, $\Omega_*$, on the luminosity.}
Luminosity normalized by $L_{0}$ is shown for different values of $\Omega_* / \Omega_o$. The dashed black curve represents a simple superposition of the orbital and spin contributions, while the red solid line is a fitting of the numerical values, capturing the asymmetry due to the aligned/anti-aligned character in angular momenta. 
}
 \label{fig:L_spin}}
\end{figure}

In order to understand the impact of the NS spin, we shall fix a set of representative parameters for the orbital motion and consider different values for $\Omega_*$. In particular, we pick up an orbit with frequency $\Omega_o = 0.25 c/R_o$  and separation $R_o = 2.5 R_*$, and set the spin to $\Omega_* = \kappa \, \Omega_o $, varying $\kappa$ from $-1$ to $1$. Note that choosing negative values of $\kappa$ corresponds to anti-aligned orbital and spin angular momenta.  
We measure the luminosity and normalize it with $L_{\rm orb}$, which corresponds to its value for a purely circular orbit at angular velocity $\Omega_o$ (i.e., $\chi=0$, $\Omega_* = 0$).
Figure~\ref{fig:L_spin} shows the results, together with a curve (dashed black) that represents an estimation from the simple addition of the orbital and spin contributions (i.e. $L_{\rm orb} + L_{\rm spin}$); and a fitting (red solid line) of the form,
\begin{equation}
 L \approx c^{-5}\mu^2 \Omega_{o}^4 \left[  w_0 + w_1 \kappa + w_2 \kappa^2 + w_3 \kappa^3 + \kappa^4 \right] ,
\end{equation}
where $ w_0 = \frac{4}{15} (v_{o}/c)^2 f(v_{o}/c) \approx 0.042$ comes from the pure orbital part, while the last term represents the pure spin contribution. 
The other coefficients are fitted from the numerical data, giving: $w_1 \sim 0.06$, $ w_2 \sim 0.35$ and $w_3 \sim 0.23 $. They account for the non-trivial superposition of the two dynamical effects.
We note that for the case in which the spin and orbital angular momenta are anti-aligned the resulting luminosity is very close to the direct sum of each contribution, whereas for aligned cases there is an extra enhancement.  
Also, it is worth mentioning that the orbital motion produces typically much weaker luminosity than the NS spin alone (i.e., pulsars); for this particular setting being $\sim5$\% of the pulsar spin-down luminosity. The reason for this is that the luminosity by the orbital motion is $(v_o / c)^2$ smaller than that by the spin motion from the post-Newtonian viewpoint.  On the other hand, we also notice that when the orbit and spin are synchronized, then the pulsar luminosity can be significantly enhanced (for this case, on about $70$\%).

\begin{figure*}
	\centering
\includegraphics[scale=0.163]{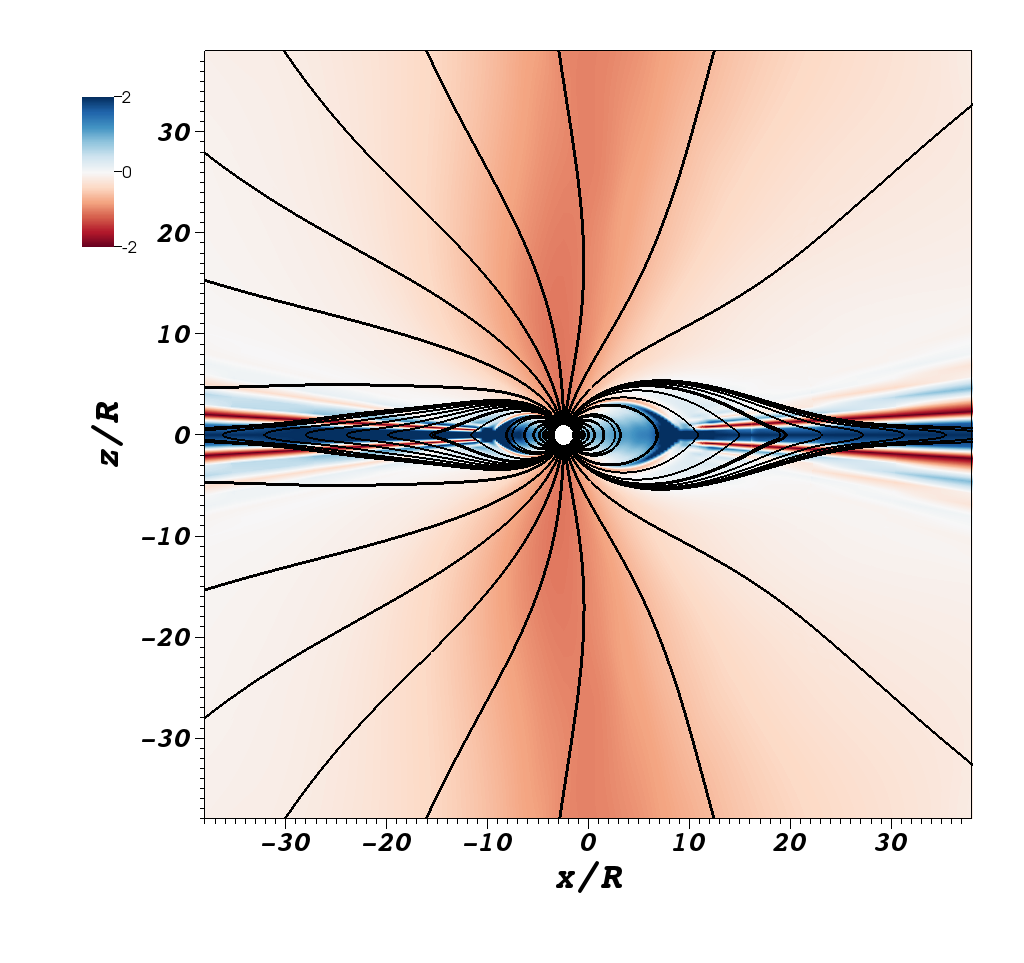}
\includegraphics[scale=0.163]{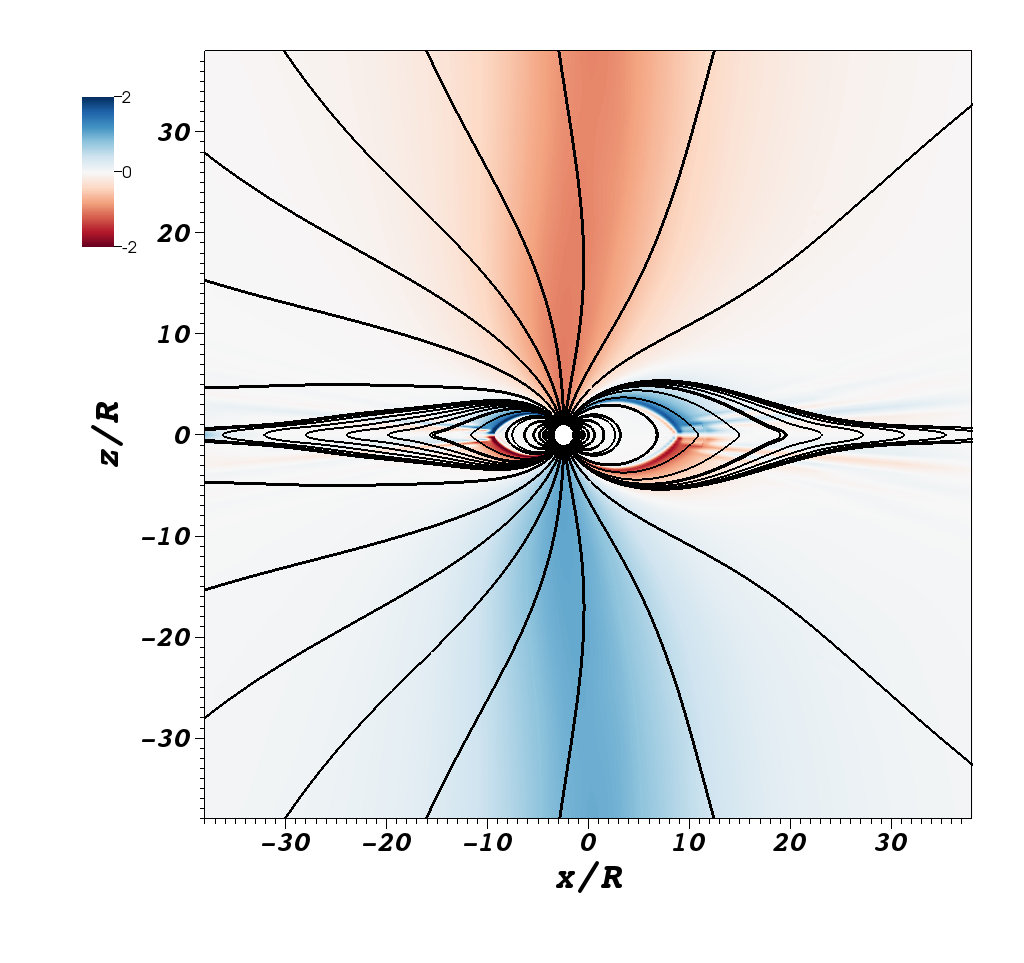}
\includegraphics[scale=0.163]{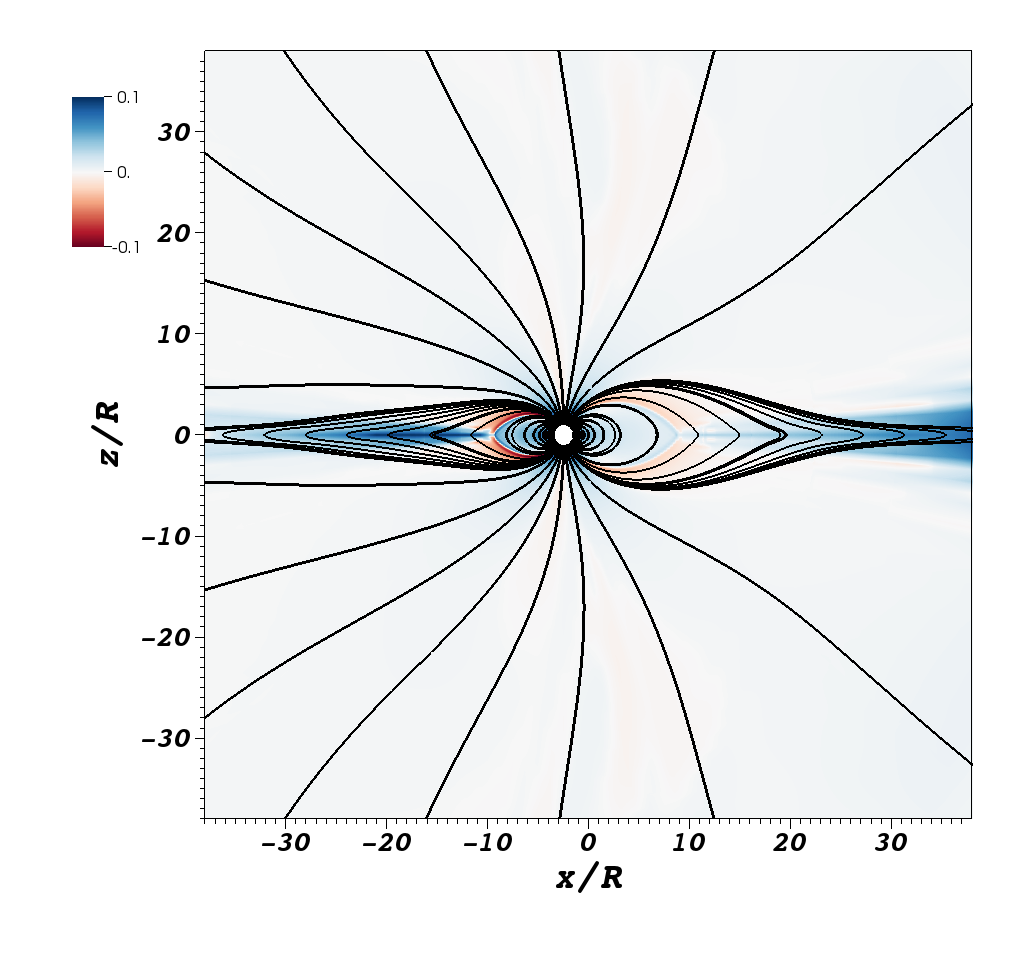}

  \caption{\textit{Several quantities for synchronized orbital motion, with $\Omega_o = \Omega_* = 0.25 c/R_o$ and $R_o = 2.5 R_*$.} 
  Stationary solution after $t=2.5$ periods, on the $y=0$ plane.  Charge density distribution (left panel) and electric currents along magnetic lines (middle panel) are normalized by $\Omega_o B / 2\pi c$ and  $\Omega_o B / 2\pi$, respectively. Right panel: Lorentz invariant parameter $\varrho$ from Eq.~\eqref{eq:currents} (multiplied by $r^2$ to improve the contrast). Black lines show some representative magnetic field lines along the plane.
  }
	\label{fig:co-plane_sync}
\end{figure*}

In Fig.~\ref{fig:co-plane_sync} we display the charge and current distributions --like the previous plots in Fig.~\ref{fig:co-plane}--, for the case in which the spin is synchronized with the orbit ($\Omega_* = \Omega_o $).
As seen before, for the synchronized motion, the spin effect tends to ``dominate'' over the orbital one, thus resulting in a magnetospheric configuration very similar to that of pulsars.
Both electric charge and parallel current distributions are consistent with typical aligned pulsar results (see, e.g.,~\cite{bai2010, parfrey2012}): 
strong current layers manifest as a local enhancement of the parallel current components $j_{\parallel}$ (middle panel), with further currents (of opposite signs for each hemisphere) flowing along the poloidal field lines.     
Although their intensities are approximately twice stronger than those in the pure spinning setting, in line with the luminosity being about twice larger as well. 
Another significant difference is again the location of the Y-point at each side of the NS, which we find to be approximately given by, $\hat{r} \approx c/\Omega_o \pm R_o$.
Spacelike currents are present (in red color, right panel) along the current layers, surrounding the equatorial CS and also within a small region inside the polar cap. 

\subsection{Misalignment effects}

\begin{figure}
\centering{
\includegraphics[scale=0.33]{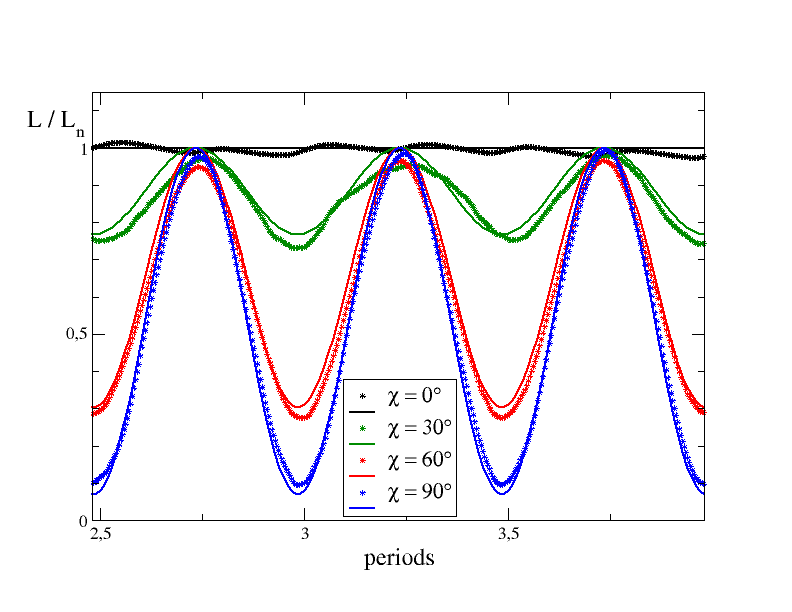}\\
\includegraphics[scale=0.33]{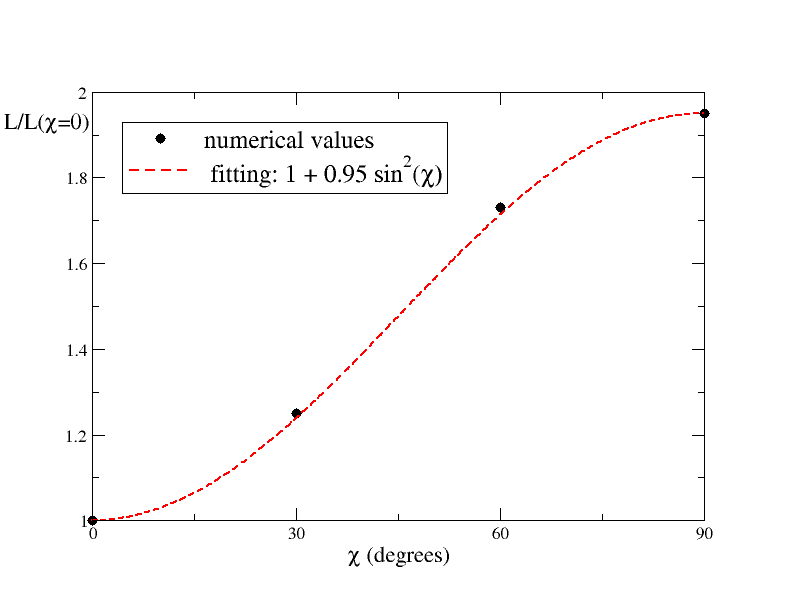}
\caption{\textit{Effects of misalignment $\chi$ on the luminosity.} The two limiting cases, $\Omega_* = 0$ and $\Omega_* = \Omega_o$, are considered. 
\textbf{Top panel ($\Omega_* = 0$):} luminosity computed at a given radius (inside the wave zone), as a function of time. 
Luminosity is normalized with, $L_n := L(\chi=0) \, \left( 1 + 0.5 \, \sin^2 \chi \right) $. 
Symbols represent the numerical values, while the solid curves corresponds to a fitting: $L/L_n \approx \left( 1 - 0.93 \, \sin^2 (\Omega_o t) \, \sin^2 \chi \right) $.
\textbf{Bottom panel ($\Omega_* = \Omega_o$):} comparison of the constant luminosity inside the light cylinder at different misalignment's, showing a pulsar behavior: $ L / L(\chi=0) \approx \left( 1 + 0.95 \, \sin^2 \chi \right) $.
}
 \label{fig:L_misaligned}}
\end{figure}

We focus now on the cases in which the magnetic axis is not aligned with the orbital angular momentum. To get the main aspects of the influence of this angle, without exploring the whole parameter space, we shall fix the 
same parameters $\Omega_o $ and $R_o$ as in Sec.~\ref{spineff}, and consider several angles $\chi = \{$ $0$\textdegree, $30$\textdegree, $60$\textdegree, $90$\textdegree $\}$ for two NS spins, $\Omega_* = \{ 0, \Omega_o \} $. These two limiting cases display very different behaviors, as shown in Fig.~\ref{fig:L_misaligned}: for the synchronized motion with $\Omega_* = \Omega_o$ (bottom panel)  the spin effect completely dominates --as expected from the previous section results--, and thus, the resulting luminosity resembles that of an oblique pulsar, i.e., $ L \approx L(\chi=0) \left( 1 + \sin^2 \chi \right) $ (see, e.g., \cite{spitkovsky2006}). 
On the other hand, in the pure orbital setting, $\Omega_* = 0$ (top panel), the luminosity at a given radius exhibits a strong phase dependency~\footnote{ Note that such phase modulations also appear in the analytic estimations for vacuum magnetospheres, arising from a term proportional to $(\vec{\mu}\cdot \dot{\vec{a}})^2$ (see the second term of Eq.~(A6) in~\cite{ioka2000}).}. Inside the region given by $\hat{r} \lesssim c/\Omega_o$, this can be modeled quite well with, $L \approx L(\chi=0) \, \left( 1 + 0.5 \, \sin^2 \chi \right) \left( 1 - 0.93 \, \sin^2 (\Omega_o t) \, \sin^2 \chi \right) $.

\begin{figure*}
	\centering
\includegraphics[scale=0.163]{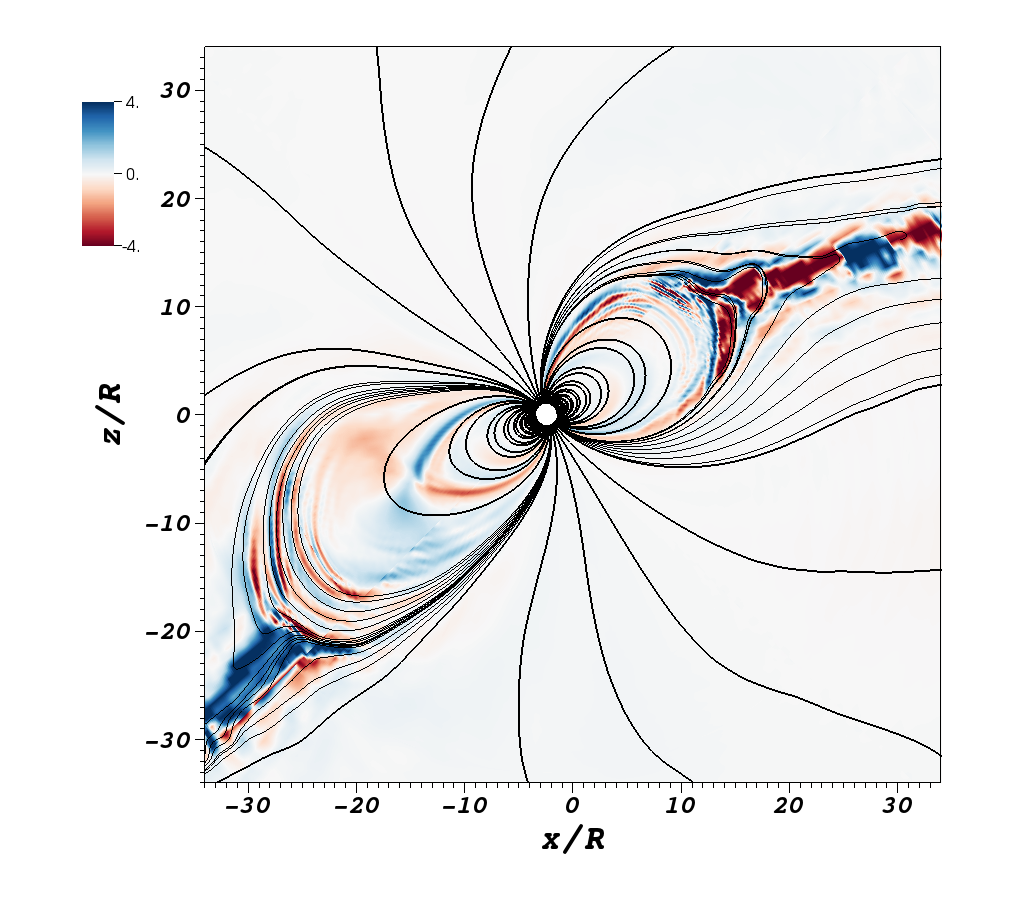}
\includegraphics[scale=0.163]{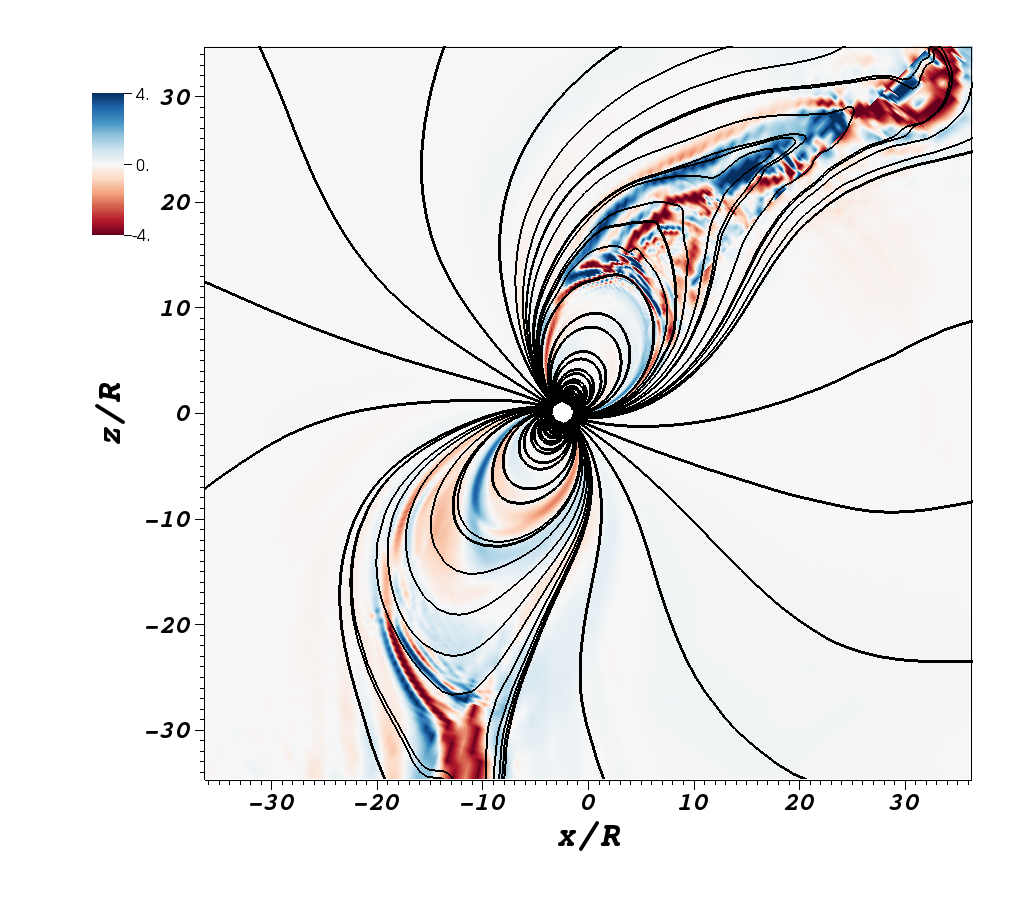}
\includegraphics[scale=0.163]{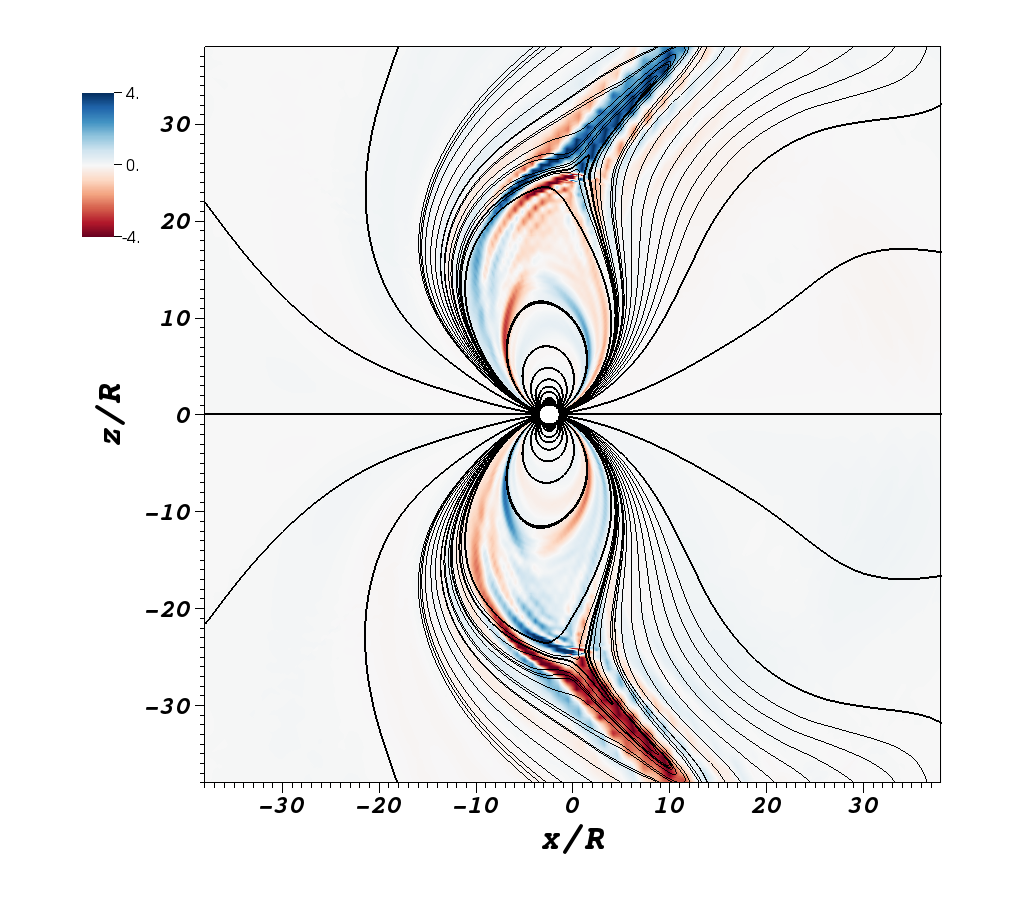}
  \caption{\textit{Effects of misalignment on the magnetosphere with $\Omega_* = 0$.} 
  Parallel electric currents, $ 2\pi j_{\parallel} / \Omega_o B $, on the $y=0$ plane after $2.5$ orbits are shown in color scale, for angles $ \chi = 30$\textdegree (left panel), $ \chi = 60$\textdegree (middle panel) and $ \chi = 90$\textdegree (right panel).
  Black lines show some representative magnetic field lines along the plane.
 }
	\label{fig:co-plane_misaligned}
\end{figure*}

Figure~\ref{fig:co-plane_misaligned} illustrates the magnetosphere with $\Omega_* = 0$, after $2.5$ orbits, for three misalignment angles. Although the configurations are intrinsically three-dimensional, some insight can be gained from considering them at the co-orbiting plane (defined by the orbital axis and the vector pointing to the NS from the CoM: in this case, the $y=0$ plane). In color, we represent the electric current component along the magnetic field, together with some representative magnetic field lines projected to the plane. Alternate patterns on the currents can be seen again inside the close zone, like in the aligned setting. Such currents become particularly intense ($j_{\parallel} \sim 4 \, \Omega_o B / 2\pi$)  at the CS, that oscillates here about the dipole equator. These shapes are reminiscent to those of oblique pulsars, although there the CS oscillate about the rotational equator instead.
Finally, we notice that for the orthogonal case (i.e., $\chi=90$\textdegree) the CS looks steady and smooth, in contrast to the intermediate inclinations for which there are signs of magnetic reconnection activity and plasmoids~\footnote{This fact should be taken with some caution, since it may be indicating lack of numerical resolution/dissipation.}. 

\subsection{Inspiral Orbits}

In this section, we analyze the evolution of the luminosity during the late stages of an inspiral orbit.  
Even though in this paper we are not including curvature effects, we shall mimic the inspiral phase of a binary system by taking relevant astrophysical parameters to set the NS trajectory around the CoM. 
To that end, we consider similar parameters to those used in \cite{paschalidis2013}. Specifically, we take the mass ratio of the BHNS binary to be $q=M_{\rm BH}/M_{\rm NS} = 3$, with $M_{\rm NS} = 1.4 M_{\odot}$ and stellar compactness, $\mathcal{C} \equiv G M_{\rm NS} / c^2 R_*  = 0.2$. 
The total mass is thus given by $M= 5.6 M_{\odot}$. Instead, for the BNS trajectory we take an equal mass system, i.e., $q=1$, and hence, $M= 2.8 M_{\odot}$.  Assuming quasi-circular inspiral orbits, the angular frequency evolves approximately according to~\cite{ShapiroTeukolsky}, 
\begin{equation}
\Omega_{o}(t) = \left[ \Omega_{o,0}^{-8/3} - \frac{256}{5} \frac{G^3}{c^5} \frac{q M^{5/3}}{(1+q)^2} (t-t_0 ) \right]^{-3/8}  ,
\end{equation}
with $\Omega_{o,0}$ being the initial orbital frequency at $t=t_0$. Our orbital radius then reads,
\begin{equation}
 R_o (t) =  \frac{q}{1+q} \left(  \frac{G M}{(\Omega_{o}(t))^2}\right)^{1/3} ,
\end{equation}
where we used $ \Omega_o = \displaystyle\sqrt{\frac{G M}{a^3}}$ with the orbital separation, $a$, through $R_o \equiv \displaystyle\frac{q}{1+q} a$.
We start from a quasi-stationary initial configuration with $R_{o,0} \simeq 5.5 R_* $ (and its corresponding $\Omega_{o,0}$) at $t=t_0$, following the inspiral evolution from that moment on, until reaching a final radius $R_{o} \sim 2 R_* $. The expected disruption radius in our setting would be $R_{o}^{\rm dis} \simeq 2.4 R_* $, as estimated \cite{paschalidis2017} from $a^{\rm dis} \simeq 3 M_{\rm BH} \left( \mathcal{C}/0.2\right)^{-1} \left(  q/7\right)^{-2/3} $.

\begin{figure}
\centering{
\includegraphics[scale=0.33]{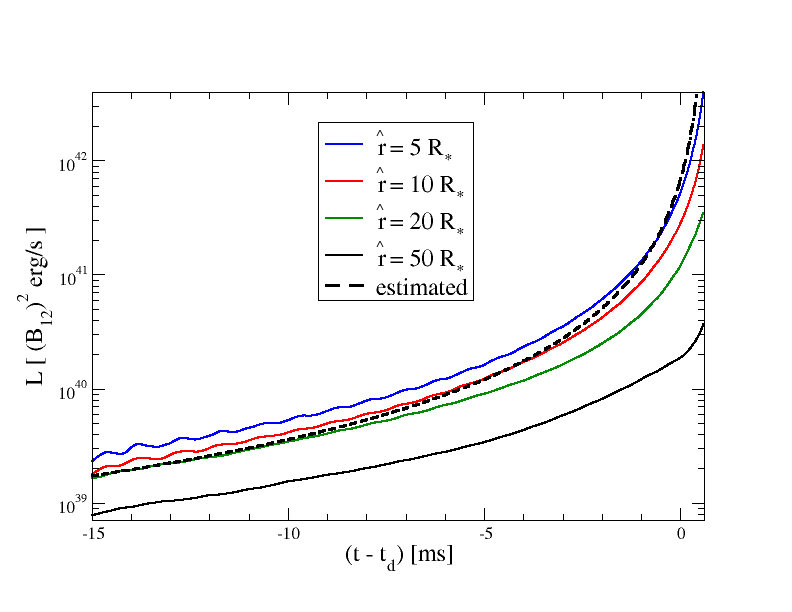}\\
\includegraphics[scale=0.33]{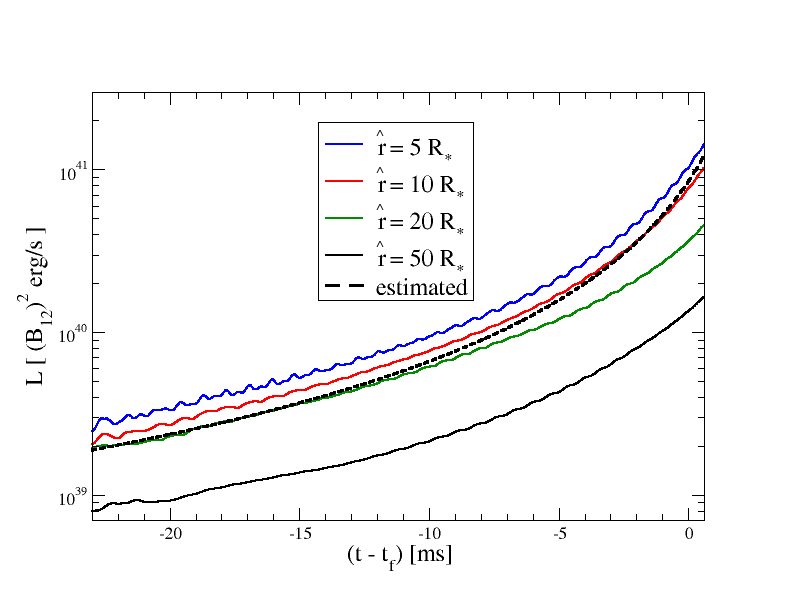}
\caption{\textit{Evolution of the luminosity for inspiral quasi-circular orbits along two inspiral trajectories}. The luminosity is measured at representative spheres enclosing the NS, at radius $\hat{r} = \{ 5, 10, 20, 50 \} R_* $.
The parameters defining the orbits are a mass ratio $q=3$ and total mass $M= 5.6 M_{\odot}$ for the BHNS trajectory (top panel); and $q=1$ with $M= 2.8 M_{\odot}$ for the BNS case (bottom panel).
}
 \label{fig:L_inspiral}}
\end{figure}

The evolution of the luminosity during the trajectory for several extraction radii is shown in Fig.~\ref{fig:L_inspiral} (top panel), where time is presented in physical units and relative to $t_{\rm d}$; the time for which the estimated disruption radius $R_{o}^{\rm dis}$ is attained. The luminosity is given in physical units, as well, for a typical magnetic field strength at the stellar pole of $10^{12}$\,G. Since luminosity scales exactly as $B^2$ in force-free electrodynamics, their values can be re-scaled by $B_{12}^2 \equiv \displaystyle \left(\frac{B}{10^{12}\,{\rm G}}\right)^2 $, as indicated in the plot.  
From the initiation of the inspiral motion, the luminosity is no longer constant inside the light cylinder $\sim c/\Omega_o (t)$, since the rising of the luminosity initiates at the star and it takes some time to propagate outwards. This is clearly seen from different curves representing integration of the flux at several spheres enclosing the NS.
Note that the luminosity measured at $\hat{r} = 50 R_*$ departs from the other values from the very beginning, reflecting the fact that a significant fraction of the Poynting flux gets dissipated at the CS.
Such difference is further enhanced later, reaching $\sim 2$ orders of magnitude (with respect to the value at $\hat{r}=5R_*$), by the time-delay effect already mentioned.
The extra (black dashed) curve, included in the plot, depicts the predicted/estimated values from our previous results of stationary circular orbits (see Eq.~\eqref{eq:ffe}). It can be noted that such estimated curves match with the dynamical results quite well (especially when the comparison is made taking the values at the integration radius close to $\sim c/\Omega_o (t)$), suggesting that the time-scale of the magnetospheric response is comparable to the inspiral one during the whole evolution.    

\begin{figure*}
	\centering
\includegraphics[scale=0.163]{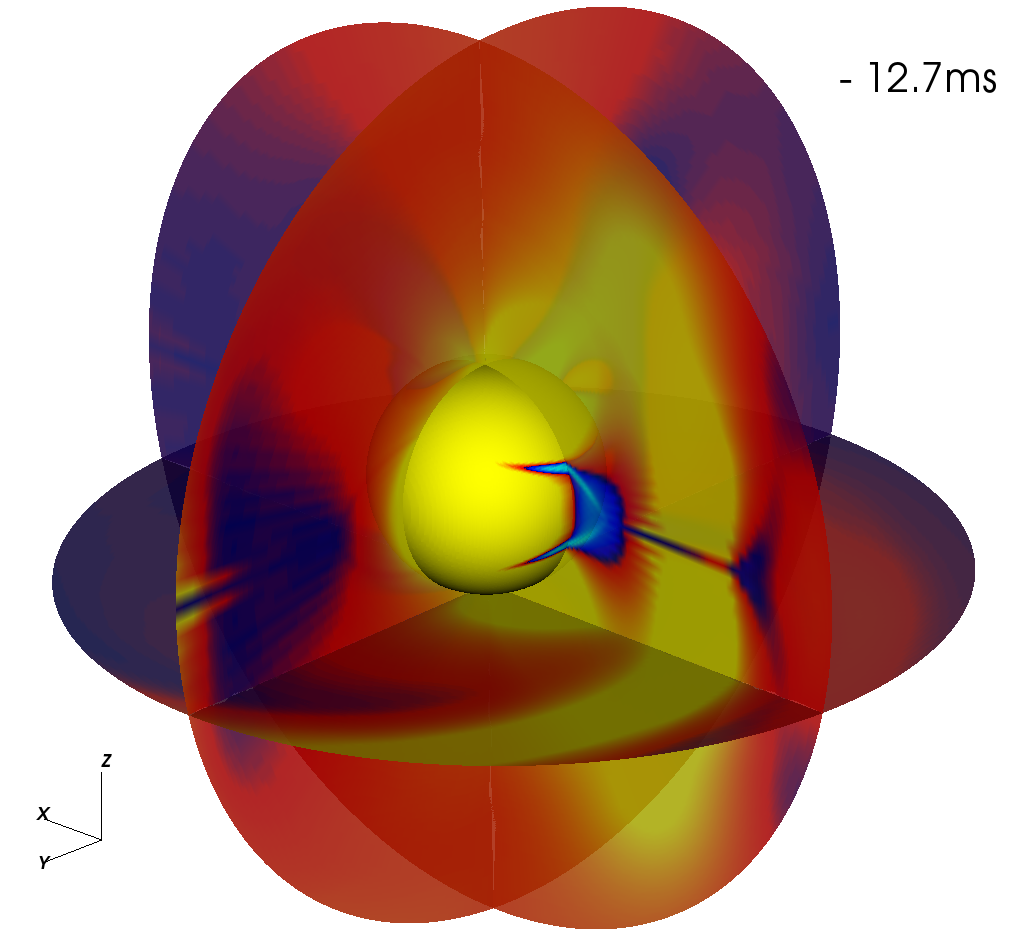}
\includegraphics[scale=0.163]{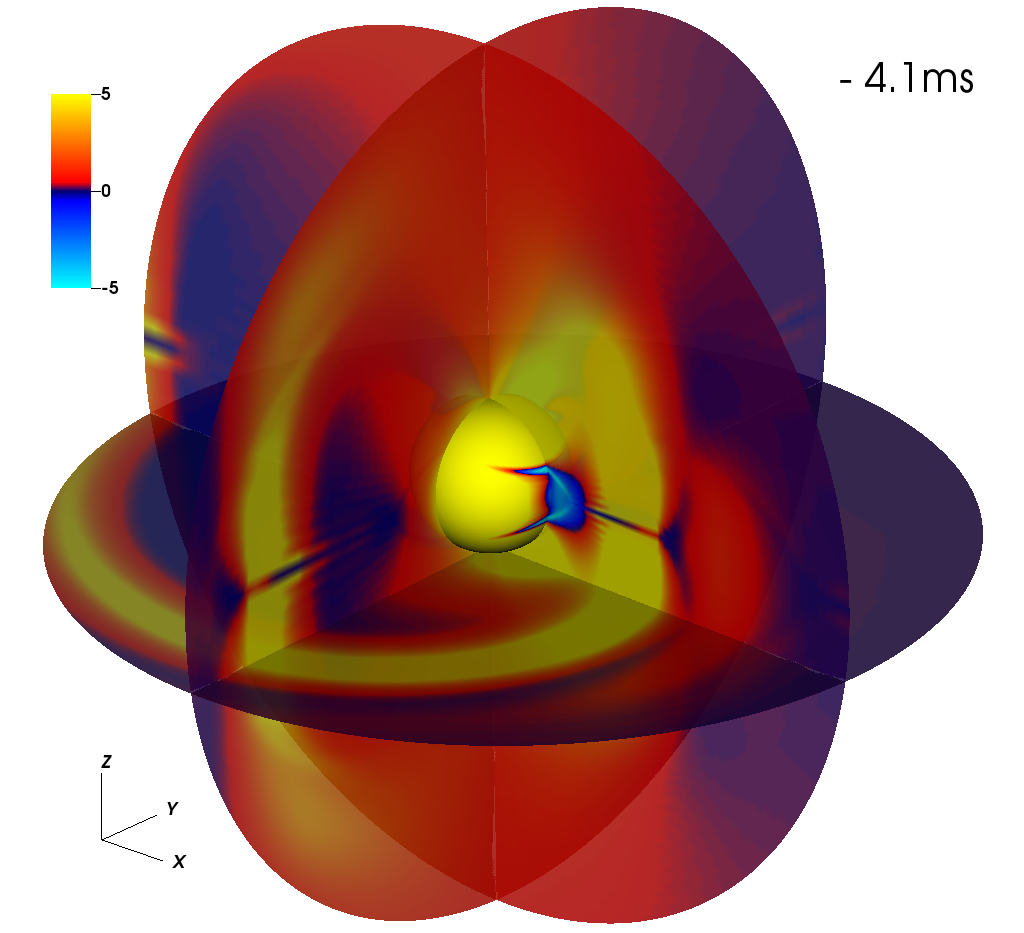}
\includegraphics[scale=0.163]{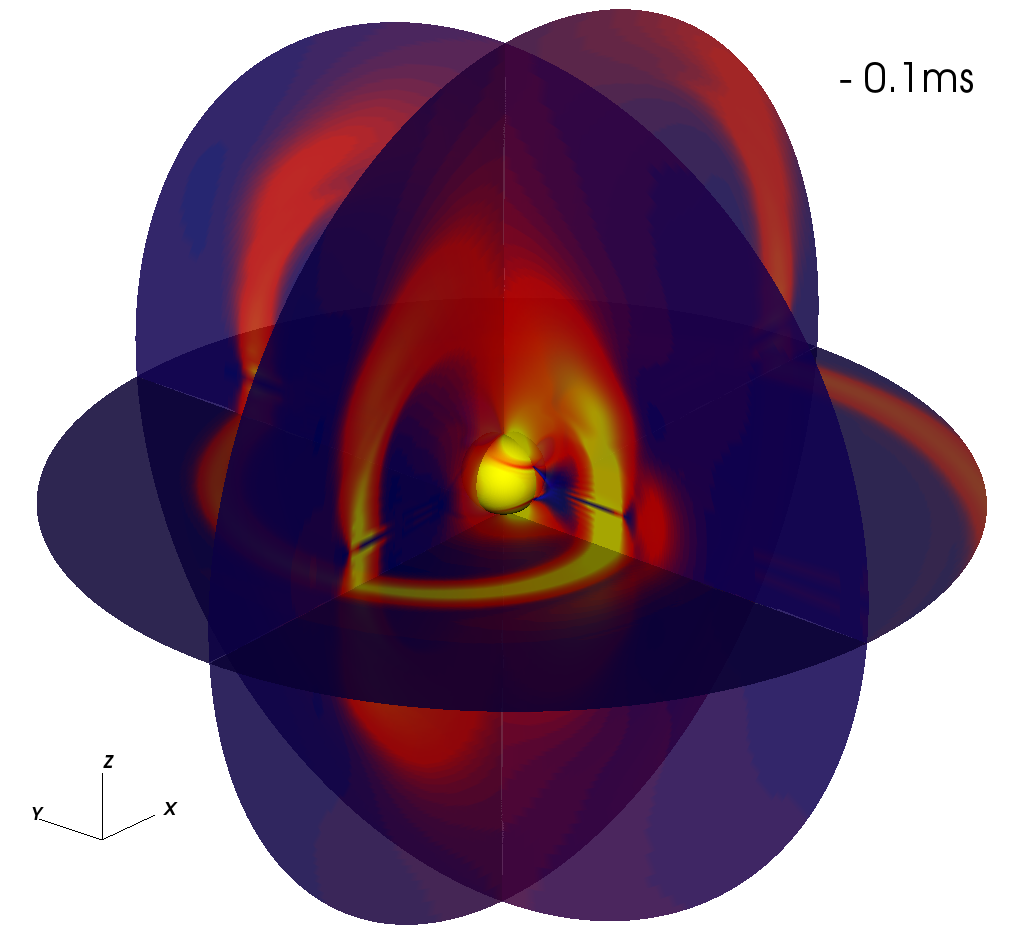}

  \caption{\textit{Poynting flux for inspiral quasi-circular orbit of a BHNS binary with mass ratio $q=3$ and total mass $M= 5.6 M_{\odot}$.} 
 Snapshots of the Poynting flux normalized by $L_0 (t) / 4\pi r^2$ at three stages, $t-t_d = \{-12.7, -4.1, -0.1 \}$ ms, during the inspiral trajectory are presented. Projections on several slices (i.e., $x=0$, $y=0$, $z=-c/\Omega_{o}(t)$ and $r= c/\Omega_{o}(t) $) are displayed, to illustrate its three-dimensional distribution. The state of the system at each snapshot can be described as follows: $a \sim 6.6 R_*$, $v_o \sim 0.26 c$ and $c/\Omega_o \sim 19 R_*$ (left panel);  $a \sim 5.1 R_*$, $v_o \sim 0.3 c$ and $c/\Omega_o \sim 12.9 R_*$ (middle panel);  $a \sim 3.3 R_*$, $v_o \sim 0.37 c$ and $c/\Omega_o \sim 6.7 R_*$ (right panel).
 }
	\label{fig:PF}
\end{figure*}

In Fig.~\ref{fig:PF} we illustrate the three-dimensional (radial) Poynting flux density along the inspiral BHNS trajectory. The values have been normalized with $L_0 (t) /4\pi r^2$, for better visualization and to facilitate the comparison among the different snapshots. 
We find that --as in the stationary circular orbits-- the flux distribution through a sphere of radius $r= c/\Omega_{o}(t)$ remains broadly unchanged, only rotating at the orbital frequency $\Omega_o (t)$. The outgoing fluxes outside the ``instantaneous light cylinder'', i.e., $c/\Omega_{o}(t)$, forms spiral structures over the $x$-$y$ plane(s), producing the equatorial CS (with magnetic reconnections) as these Alfven fronts propagate outwards.

We now estimate how the inclusion of curvature (associated to both the NS and the BH) can impact on the luminosity. To that end, we shall pick up a state at an orbital separation of $a\approx 6.6 R_*$ (represented in the middle panel of Fig.~\ref{fig:PF}) and compare its luminosity with the results reported in \cite{paschalidis2013} for essentially the same parameters. Our results show $L \approx 2.4 \times 10^{39} \, B_{12}^2 \, {\rm erg/s}$, which should be associated in our case exclusively to the MD acceleration mechanism.
On the other hand, they obtained $L\approx 6.2 \times 10^{40} \, B_{12}^2 \, {\rm erg/s}$, for the case where both compact objects were non-spinning. Hence, taken at face value, one might say that the luminosity is enhanced by a factor $\sim 25$ due to curvature effects. However, a few important considerations are necessary. First, in \cite{paschalidis2013} the luminosity is measured at large distances, $r\gtrsim 90 R_*$, and not inside the light cylinder where we compute it. As we found dissipation at CS outside the light cylinder, the comparison is not direct, and perhaps even slightly larger luminosity could be expected at the light cylinder when curvature is accounted for. 
On the other hand, it is not very clear how to disentangle from different curvature effects like unipolar induction, estimated in this context to give $L_{UI} \approx 6.4 \times 10^{39} \, B_{12}^2 \, {\rm erg/s}$ \cite{paschalidis2013}. 

Similarly to the BHNS binary scenario, we also follow the evolution of the luminosity for an equal mass BNS system (see bottom panel of Fig.~\ref{fig:L_inspiral}). The trajectory is evolved from an orbital separation $a_i \sim 6 R_*$ up to a final one $a_f \sim 3 R_*$ (which sets the reference time $t_f$).
Thus, we consider a complementary dynamical range with respect to the one studied in \cite{palenzuela2013electromagnetic, palenzuela2013linking, ponce2014}, where the simulations begins at about our final orbital separation and proceeds though the late inspiral phase until (and after) the merger. 
Although a comparison of the luminosity here is not as direct as before, we can estimate from their case ``$U/u$'' of a weakly magnetized companion, $L\approx 10^{42} \, B_{12}^2 \, {\rm erg/s}$ at orbital separation $a\sim 3 R_*$, 
whereas we obtain $L \approx 10^{41} \, B_{12}^2 \, {\rm erg/s}$.  Therefore, the enhancement attributed to the effects of curvature is again approximately an order of magnitude.

\subsection{Implication to observation}

Here, we briefly discuss the implications of our numerical results. As shown in Sec.~III A, CS are always developed in the local wave zone, near the radius $c/\Omega_o$, for orbiting NSs. This feature is shared with the pulsar magnetosphere, for which CS are also developed just outside the light cylinder at radius $\agt c/\Omega_*$. This suggests that some EM signals similar to the ones in pulsars are likely to be emitted from the orbiting NSs, even in the absence of NS spin. In the following, we estimate the luminosity of the EM signals referring to the latest studies of magnetospheric emissions from pulsars. 

Before going ahead, we show approximately how much EM energy can be emitted in total during the inspiral phase. For this estimate, we simply employ Eq.~\eqref{eq:zeroth} with the orbital evolution determined by the Newtonian gravity plus gravitational radiation reaction via quadrupole formula as (e.g.,~\cite{ShapiroTeukolsky})
\begin{eqnarray}
\dot a=-{64G^3M^3 \eta\over 5 a^3 c^5},
\end{eqnarray}
where $\eta$ is the symmetric mass ratio, i.e., the ratio of the reduced mass to the total mass, $M$, of a binary and written as $q/(1+q)^2$. Then, the maximum total energy dissipated in EM waves is calculated as,
\begin{eqnarray}
\int_{t_i}^{t_f} L_0 dt &=& -\int_{a_i}^{a_f}L_0 {5 a^3 c^5 \over 64G^3M^3\eta} da \nonumber \\
&=& {B^2R_{*}^3 q \over 64} \left({R_{*} \over a_f}\right)^3 \nonumber \\
&\approx&2 \times 10^{39}\,{\rm erg}
\left({B \over 10^{12}\,{\rm G}}\right)^2
\left({R_* \over 12\,{\rm km}}\right)^6
\nonumber \\
&&~~~~~~~~~~\times 
\left({a_f \over 42\,{\rm km}}\right)^{-3}
\left({q \over 3}\right), \label{eq180}
\end{eqnarray}
where $B$ is the magnetic-field strength at the pole (i.e., $B=\mu/2R_*^3$), $a_i$ and $a_f$ denote the initial orbital radius at $t=t_i$ 
and the orbital radius at the onset of merger at $t=t_f$, respectively, 
and we assume that $a_i \gg a_f$. In this equation, 
we suppose that $M_{\rm BH} \approx 4.2 M_\odot$, 
$M_{\rm NS} \approx 1.4M_\odot$, and $a_f \approx 5M$. As we have shown in this paper, the luminosity is enhanced in the close binary orbits, and hence, the total energy emitted could be larger by a factor of several than Eq.~(\ref{eq180}). Nevertheless, 
Eq.~\eqref{eq180} indicates that for the typical magnetic field strength at the NS pole of $B=10^{12}$\,G, the available energy for the EM signals is at most $\sim 10^{40}$\,erg, and thus, for the observation being possible, the presence of an efficient emission mechanism or unusually high magnetic field strength would be necessary. 

The latest PIC simulations (e.g.,~\cite{philippov2018}) show that near the so-called Y-point at the light cylinder of the pulsar magnetosphere, CS are developed, and as a result, the reconnection of magnetic field lines is enhanced. In such a region, the magnetic field strength and number density of the electron-positron pair are significantly increased. High-energy electrons and positrons accelerated in the strong magnetic field become the sources of the synchrotron and inverse Compton radiations. Because the particle energy is high, the energy can be extended to the MeV and GeV gamma-ray bands. Such an emission is consistent with the presence of the gamma-ray pulsars observed by {\em Fermi} satellite for the isolated pulsar case~\cite{Fermi2013}. The luminosity of the gamma-ray pulsars could be 1--100\% of the spin-down luminosity of the pulsars. In the present context, the spin-down luminosity should be replaced by the total luminosity of the orbiting NS, which is $L\alt 10^{42}\,{\rm erg/s}$ for typical magnetic field strength of order $10^{12}$\,G at the poles. Thus, it is reasonable to consider that the gamma-ray luminosity would be at most $\sim 10^{42}\,{\rm erg/s}$, for which the duration is $\alt 10$\,ms. 
It would be quite difficult to detect such a low-luminosity gamma-ray source with very short duration using the current and near-future gamma-ray telescopes, if we suppose that the typical distance to the source is $\agt 100$\,Mpc. This would be also the case for the observation of X-rays. 
If the magnetic field strength is as high as that of magnetars, $\sim 10^{14}$\,G at its pole, the luminosity could be enhanced and the gamma-ray signals may be observable. However, such high magnetic fields may not be very likely for inspiraling NSs, as the observational results of NSNS in our Galaxy indicates~\cite{Pol2019}.

As discussed in Sec.~III D, \cite{paschalidis2013} suggests a significant enhancement of the EM luminosity if curvature effects, i.e.,  interaction of the magnetic field with the BH, are included: at least an order of magnitude seems possible.
Although it is not clear how this extra luminosity is distributed in terms of the magnetospheric configuration, that is, how to disentangle from the unipolar induction mechanism, in which the two compact objects are connected through a twisted bundle of magnetic field lines and thus a DC-circuit established in the near zone, localized plasma winds in between the two compact objects are expected to produce curvature radiation dominating in the $\gamma$-rays band, and synchrotron emissions in the hard $X$-rays and soft $\gamma$-rays bands \cite{mcwilliams2011}. It was also suggested that a fraction of this flux within the bundle, carried in the form of plasma kinetic energy, will reach the primary NS surface and form a hot spot emitting thermal energy as $X$-rays~\cite{mcwilliams2011}.
This flux-tube UI configuration was further proposed to model fast radio bursts (FRBs) arising from coherent curvature radiation in the late-inspiral phase (but, prior to the last few orbits) in NSNS binaries \cite{wang2016}.
All these speculations encourage us to perform force-free simulations including general relativistic effects, and also to model this radiation along the lines of \cite{bai2010} to obtain light-curves from our solutions.

The latest work also shows that in the reconnection region, radio waves may be emitted by coalescence of magnetic islands in the CS that produces magnetic perturbations propagating away~\cite{uzdensky2013, philippov2019pulsar, lyubarsky2019}. For this process, the predicted spectral flux density of the radio waves with the frequency $\nu$ is
\begin{eqnarray}
S_{\nu} \sim {B_L^2 \over 8\pi}(\Gamma \ell)^3 (\pi D^2)^{-1}(\tau \nu)^{-1},
\end{eqnarray}
where $B_L$ denotes the magnetic-field strength for the reconnection region, $\Gamma$ and $\ell$ are the bulk Lorentz factor and characteristic scale of the magnetic islands, $D$ is the distance to the source, and $\tau \sim 10/\omega_p$ with $\omega_p$ the plasma frequency. Here, $\ell$ is determined by the force balance and energy conservation in the magnetic islands, and $\Gamma$ is inferred as 10--100~\cite{uzdensky2013,lyubarsky2019}. In the present context, we obtain a quite small flux for orbits close to the innermost stable circular orbit
\begin{eqnarray}
S_{\nu} &\sim &2 \times 10^{-8}\,{\rm Jy}
\left({B_L \over 4 \times 10^7 \,{\rm G}}\right)^2
\nonumber \\
&&\times 
\left(\Gamma \over 100\right)^3
\left( {\ell \over 10\,{\rm cm}}\right)^3
\left( {D \over 100\,{\rm Mpc}}\right)^{-2}. \label{eqS1}
\end{eqnarray}
Here we suppose that the total mass of the system is $5.6M_\odot$, $a \approx 100$\,km, and the magnetic field strength at the NS pole is $B=10^{12}$\,G. We also simply set $\tau \nu=1$. The expected value of $\ell$ in the reconnection region of the pulsar magnetosphere is about 10 times larger than the thickness of the CS, which is proportional to $B_L^{-3/2}$ as~\cite{uzdensky2013, lyubarsky2019} 
\begin{eqnarray}
\sim 0.1 \,{\rm cm}\left({B_L \over 10^8 \,{\rm G}}\right)^{-3/2}.
\end{eqnarray}
This is the reason why, for the strong magnetic fields in the reconnection region, the intensity of this type of the radio waves is low (i.e., in this scenario, the luminosity decreases with the decrease of the orbital separation, $a$). If the magnetic field strength for the late inspiral stage of NS binaries is 
$B \approx 10^{10}$\,G, the luminosity would be $10^5$ times lager than that of Eq.~(\ref{eqS1}) because the width of the magnetic islands becomes larger. However, the flux is still $\sim 1$\,mJy. 

Since the size of each magnetic island is likely to be quite small for the strong magnetic field case, a large number of the magnetic islands may be simultaneously generated. As discussed in Ref.~\cite{lyubarsky2019}, fast waves from many merging magnetic islands may nonlinearly interact and transfer the energy into the plasma. If this happens and the emission occurs in an optimistically coherent way, the predicted luminosity is written as ~\cite{lyubarsky2019}
\begin{eqnarray}
L_n &\sim& {\Omega_o \over 2\pi \nu}L 
=3 \times 10^{35}\,{\rm erg/s}
\left(\Omega_o \over 2000\,{\rm rad/s}\right)
\nonumber \\ 
&& ~~~~~~~~~~~~~~~\times
\left( \nu\over 1\,{\rm GHz}\right)^{-1}
\left( L \over 10^{42}\,{\rm erg/s}\right),
\end{eqnarray}
which approximately leads to the following flux density,
\begin{eqnarray}
S_{\nu} &\sim & 0.1\,{\rm mJy}
\left(\Omega_o \over 2000\,{\rm rad/s}\right)
\left(\nu \over 1\,{\rm GHz}\right)^{-2}
\nonumber \\
&&\times 
\left(L \over 10^{42}\,{\rm erg/s}\right)
\left({D \over 100\,{\rm Mpc}}\right)^{-2}.
\end{eqnarray}
Thus, a much higher flux than Eq.~\eqref{eqS1} can be predicted, although the flux is still low for the detection by the current wide-field-of-view radio telescopes like CHIME~\cite{CHIME} and OVRO-LWA~\cite{OVRO}. We note that in a small fraction of pulsars~\cite{Manchester2005}, a substantial fraction of the dissipation energy of order $10^{-4}$ is emitted in the radio band. If the orbiting NSs can have such a high efficiency for the radio emission, the radio waves may be observable. 

As indicated above, it will not be very easy to detect an EM precursor of NS mergers. However, the following point should be kept in mind: In the absence of tidal disruption in the system of BHNS binaries, no EM counterparts are expected {\em after} the merger. Even in such cases, a precursor associated with the moving NS with magnetic fields can be emitted as an EM counterpart of the merger of BHNS binaries. Distinguishing a BHNS binary with a fairly large chirp mass from binary black holes is not an easy task only in the detection of gravitational waves, because the gravitational waveforms for two cases are quite similar. The precursor for BHNS binaries is likely to be emitted always if the magnetic field strength of the NS in binaries is as strong as that for the typical isolated NSs. The observation of the precursor for the BHNS binaries will play an important role for surely identifying the BHNS binaries.

It is also worth mentioning that if a precursor is detected for NS binaries, the magnetic field strength of the NS could be estimated. This will provide us important information for the evolution of the magnetic fields in an old NS that has not experienced significant mass accretion. 


\section{Conclusions}
 
In this paper we have considered force-free magnetosphere induced by an NS orbiting in a binary system, aiming at capturing the EM effects produced by the orbital motion about the CoM of the binary (BHNS or NSNS) system; in a sense, isolating the role played by the acceleration of the MD moment of the NS, from the effects of curvature. The inclusion of curvature --more specifically a BH companion-- has been deferred for a subsequent work.
Our present approach, however, has allowed a detailed and systematic study of the magnetospheric properties of these systems in close analogy to pulsars. In particular, the existence of strong return current layers and CS enabled us to connect with known EM emission mechanisms from pulsar theory.

Before considering realistic inspiral orbits associated with the binaries, we first analyzed the properties of circular orbits; noticing --as later confirmed-- that the stationary configurations attained here are reliable approximations of the system's states throughout the inspiral. The results can be summarized as follows. For the aligned and non-spinning scenario, a strong spiral CS develops along the orbital plane, beginning at $\sim c/\Omega_o$. An alternate pattern of charges/currents is found in the near zone of the NS, enclosed by thin return current layers that form an Y-point with the equatorial CS. 
In contrast to pulsars, there is almost no charge flowing over the polar region of the NS. 
The luminosity can be well described by, $L \approx \displaystyle\frac{4}{15c^5} \mu^2 R_{o}^2 \Omega_{o}^6 f(v_o /c ) $, where $f \sim (1$--$7)$ represents relativistic corrections in $v_o /c$, to the analytic formula \eqref{eq:zeroth}. 
The orbital effect on the luminosity is much weaker than the one produced by the NS spin (comparing at a given angular frequency), although it can produce a considerable enhancement (factor $\sim 2$) on the pulsar spin-down luminosity if the motion is synchronized.
Finally, for the case that the magnetic and orbital axis are not aligned (i.e.,$\chi\neq0$), the magnetosphere resembles that of oblique pulsars, with a strong CS fluctuating about the dipole-equator (instead of the rotational equator, as in misaligned pulsars \cite{spitkovsky2006}). Interestingly, in the non-spinning case, the total luminosity acquires an orbital-phase dependency. 

We consider inspiral binary trajectories using relevant parameters for both BHNS and NSNS scenarios, finding that the evolution of the luminosity follows quasi-adiabatically from our previous estimations based on the circular orbits configurations. The Poynting fluxes emanating from the orbiting NS forms spiral structures orthogonal to the orbital axis, and produce magnetic reconnections inside the CS as these Alfven fronts propagate outwards. 
Overall, the radial Poynting flux distribution in the wave zone establishes a lighthouse effect, rotating at orbital frequency $\Omega_o (t)$ and peaking around the orbital plane; 
which is in qualitative agreement with previous GRFF studies for the cases where the companion to the primary NS is either a weakly magnetized NS \cite{palenzuela2013electromagnetic, palenzuela2013linking} or a BH \cite{paschalidis2013}.
This suggests that the EM flows beyond the light cylinder $\sim c/\Omega_o (t)$ are dominated by the MD effect, while UI occurs almost exclusively between the two compact objects in the near region.
One might interpret these two mechanisms as constituting two approximately independent sources of EM energy for the plasma (albeit both mined from the kinetic energy of the orbital motion).
However, by the interaction of the magnetic field with the curvature of the companion (if curvature is included), the field strength can be amplified by further twisting, resulting in enhanced luminosity of the MD effect --valued here to represent one order of magnitude--.

Furthermore, the contribution from each of these two mechanisms (i.e., MD/UI) will be channeled into separate processes within the magnetosphere, which then derive in the actual EM emissions.
In particular, we have devoted our attention to the existence of strong CS induced by the NS's orbital motion alone. 
These CS were not reported before: in the BHNS scenario of \cite{paschalidis2013}, this may be due to the fact that reflection symmetry about the orbital plane was imposed in their simulations;
whereas for the weakly magnetized NS companion case studied in \cite{palenzuela2013electromagnetic, palenzuela2013linking}, we see evidence of such CS forming (last panel of Fig.~10 in~\cite{palenzuela2013linking}), although it was omitted in their discussions~\footnote{Only a trailing dissipation region, making a CS tail behind the weakly magnetized NS companion, was described in these works.}.   
We argue these CS are very important due to their role in explaining radio emissions in pulsars, where recent studies~\cite{uzdensky2013, philippov2019pulsar, lyubarsky2019} have shown that the coalescence of magnetic islands (or plasmoids) in the reconnection regions can produce radio waves, as a coherent superposition of many such individual pulses.  
We have borrowed these ideas and applied them to get concrete estimations of spectral flux densities of radio waves in our context.
The predicted fluxes are not large enough as to be detected by the current wide-field-of-view radio telescopes like CHIME~\cite{CHIME} and OVRO-LWA~\cite{OVRO}. 
We conclude that only if a substantial fraction of the luminosity (e.g., $\gtrsim 10^{-4}$) is emitted in the radio band, as in some pulsars~\cite{Manchester2005}, then radio waves may be observable.


\section{Acknowledgments}

We would like to thank Kenta Hotokezaka, Kunihito Ioka, Kotha Murase, Carlos Palenzuela, Oscar Reula, Anatoly Spitkovsky, and Tomoki Wada for several helpful discussions during the realization of this work. Numerical computations were performed on a cluster in Max Planck Institute for Gravitational Physics at Potsdam.
This work was in part supported by Grants-in-Aid for Scientific Research (No. 16H02183) of Japanese MEXT/JSPS. 



\bibliographystyle{unsrt} 
\bibliography{FFE}


\end{document}